\PassOptionsToPackage{table,xcdraw}{xcolor}

\documentclass[twocolumn]{article}

\AtBeginDocument{%
  \providecommand\BibTeX{{%
    \normalfont B\kern-0.5em{\scshape i\kern-0.25em b}\kern-0.8em\TeX}}}

\usepackage{gensymb}
\usepackage{colortbl}
\usepackage{xcolor}

\usepackage[normalem]{ulem}

\usepackage{subcaption}
\usepackage{adjustbox}
\usepackage{authblk}
\usepackage{float}
\usepackage{bmpsize}

\title{Don't Look Up:  Ubiquitous Data Exfiltration Pathways in Commercial Spaces}

\author[1]{Anku Adhikari}
\author[2]{Samuel Guo}
\author[1]{Paris Smaragdis}
\author[1]{Marianne Winslett}
\affil[1]{University of Illinois at Urbana-Champaign, Illinois, USA}
\affil[2]{Carnegie Mellon University, Pennsylvania, USA}
\affil[ ]{\textit{\{aadhikr2, winslett, paris\}@illinois.edu, srguo@andrew.cmu.edu}}

\date{}
\begin{document}
\maketitle
\begin{abstract}

We show that as a side effect of building code requirements, almost all commercial buildings today are vulnerable to a novel data exfiltration attack, even if they are air-gapped and secured against traditional attacks. The new attack  uses vibrations from an inconspicuous transmitter to send data across the building’s physical infrastructure to a receiver. Our analysis and experiments with several large real-world buildings show a \textit{single-frequency} bit rate of 300K bps, which is sufficient to transmit ordinary files, real-time MP3-quality audio, or periodic high-quality still photos. The attacker can use multiple channels to transmit, for example, real-time MP4-quality video. We discuss the difficulty of detecting the attack and the viability of various potential countermeasures.

\end{abstract}

\section{Introduction}
\graphicspath{{figures/}}

Pipes are ubiquitous in buildings: fire sprinkler pipes running just above the ceiling of every office, gas pipes feeding furnaces and kitchens, water pipes supplying
restrooms, water coolers, and maintenance closets. 
In this paper, we show that in any environment where an attacker can gain access to such pipes, the attacker can attach a small, removable, unobtrusive vibrating mechanism similar to a bone-conducting headphone, and use it to transmit wide-bandwidth data to a small receiver located elsewhere in the building or at its perimeter. Whether filled with gasses, water, or air, the pipe acts as a waveguide, enabling a transmission path that can run long distances across floors and even between floors, regardless of turns and intersections, with the signal sometimes even growing louder along the way to its destination.

In this paper we provide answers to the following questions about this novel data exfiltration attack -- the {\em pipe attack}:

\begin{itemize}
    \item What kinds of pipes can be used to exfiltrate data? What kind of bitrate can be supported -- text, audio, video -- and under what conditions?
   \item Where can a pipe attack transmitter and receiver be placed in a building?  How obvious are the installation and removal process, and the installed devices themselves?
    \item How should the transmitter and receiver be designed?  What kind of signals should they transmit?  
    \item How can an attacker get the data to be exfiltrated to the transmitter and retrieve it from the transmitter?
    \item How can we detect the presence of a pipe attack?
What can be done to prevent a pipe attack or make it less effective? How expensive will these defenses be?
\end{itemize}

 In the remainder of the paper, Section \ref{sec:piperelated}  describes the state of the art for  transmission of sound through pipes. 
 Section \ref{sec:sprinklers} discusses the impact of building codes and construction cost considerations on the applicability of the pipe attack. 
 Section \ref{sec:placement}  applies this information to the problem of transmitter and receiver placement. 
 We consider the issue of getting data to the transmitter and from the receiver in Section \ref{sec:toandfrom}, and camouflaging the devices in Section \ref{sec:camouflage}.
 
Section \ref{sec:attackdesign} considers potential options for the design and implementation of a pipe attack transmitter and receiver, and describes the most effective implementation that we found for our experiments. 
  Section \ref{sec:buildings} describes the real-world building pipes and  laboratory setups  used for  experiments.
 Section \ref{sec:experimentResults} presents  measurements of the bandwidth 
 of the pipe attack in existing large buildings.
 Section \ref{sec:shannon} analyzes the empirical results to arrive at an estimated bit rate for attackers.
 
 Finally, in Section \ref{sec:defenses}, we  summarize a variety of potential defenses, none of which have proven fully satisfactory so far.

 \section{Related Research}
 \label{sec:piperelated}
 
Researchers have examined the use of  wave signals in acoustic, wireless, ultrasound, and other forms to analyse water, oil, and gas pipes and cables for  application-specific purposes  \cite{acevedo2012acoustic, mitCREATEwaterPIPEmonitoring, guided_elasticwave_PPM, oil_well-pipeline_underwateracoustics, long2003acoustic, kondis2005acoustical, hydroacoustics_undergroundpipe_soil,1555162, pipesYang16, pipeWirelessNguyen17, pipeWirelessKantaris15, pipesHe22, pipesChakraborty14, pipesChakraborty15, pipebotsDoychinov21, pipeHuang18, pipeImagesHeifetz20, pipeImagesHeifetz21, slowPipesGetreuer18, cableTrane20}. In particular, acoustic signals can be helpful in  detecting and localizing leaks and other structural health issues in water and gas pipes, and ultrasound is particularly useful for structural health monitoring of pipes and, more broadly, metal structures in general (e.g., \cite{oldHatKexel18, structuresZonzini19, structuresMalzer19, tracksYuan22}). Generally, the idea is to use the metal structure, pipe, and/or the medium inside the pipe as a signal conductor. The signals are transmitted by communicating hardware sensors  on the metal surface or inside a pipe. In the case of pipes, the pipe and medium provide a waveguide channel  for communication.  

Recent work on structural health monitoring has focused on allocating a separate channel to return the results of the monitoring, and on potential communication protocols for use on that channel. Low power usage is usually important for that application, and so reported single-frequency bit rates are low, e.g., 21 bps at 6.35W \cite{pipeWirelessKantaris15}, 100 bps at 5-10mW \cite{pipesChakraborty14, pipesChakraborty15}, and 170 bps at 224 mW \cite{structuresKexel20}. When power is not such a concern, reported rates include 100 bps \cite{ guided_elasticwave_PPM}, 250 bps \cite{pipesHe22}, 470 bps \cite{cableTrane20},  2K-10K bps \cite{pipeImagesHeifetz20,pipeImagesHeifetz21}, and 20K-40K bps  \cite{pipeHuang18}. All of this work  typically involves laboratory experiments with  pipes a meter or two long and very few if any fittings, intersections, and turns. (The exception that we are familiar with used a .9 mile 24 inch pipeline \cite{gas_pipeline_paper}, which is also very different from building pipes.) These differences make it hard to confidently generalize from  the prior work to  what bit rate we may expect from pipes in large office buildings.

Existing theoretical models of signal propagation through pipes (e.g., 
\cite{hydroacoustics_undergroundpipe_soil, kondis2005acoustical}) demonstrate that it is extremely complex to model the physics of acoustic transmission in water pipes  once fittings, intersections, and turns are taken into consideration. This strengthened our resolve  to take an empirical approach to the problem.

\section{Building Codes and Fire Sprinkler Systems}\label{sec:sprinklers}

\subsection{Building Codes}
Fire safety is a key focus of building codes worldwide.  Under the \textit{prescriptive} approach, the building code includes a set of rules that suffice to guarantee certain safety provisions in the event of a fire, based on empirical evidence from past fires and from research results. 
 
Often both prescriptive and performance-based approaches are allowed in a particular  jurisdiction.

Prescriptive building codes generally require fire sprinkler systems in commercial buildings. Performance-based codes, such as the pan-European Eurocodes \cite{eurocode}, do not.  Nonetheless, sprinkler systems greatly reduce the chance that a fire will spread enough to overcome people before they can escape, and that a fire will cause structural collapse. Thus  the \textit{voluntary} inclusion of a sprinkler system in a building design will tend to greatly reduce the expense of the structural parts of the building, the expense of materials used to protect the structural members from fire, and the size of the structural members and their fireproofing. 

The discussion that follows is based on the provisions of the prescriptive codes that require sprinklers, as they serve as a baseline for what to expect.

Almost all jurisdictions in the US and Canada have adopted a version of the National Fire Protection Association Standard for the Installation of Sprinkler Systems (NFPA 13) \cite{NFPA_13}. This standard is referenced in  the broader International Fire Code (IFC) \cite{IFC_2018} and in the International Building Code (IBC) \cite{IBC_2018}. The IBC and its sister standard for one- and two-family homes, the International Residential Code (IRC) \cite{irc_2018}, require fire sprinkler systems in almost all new construction and remodeling over a certain size, both commercial and residential. 

Jurisdictions that adopt the IBC, IRC, IFC, and NFPA 13 often add local amendments that can weaken or strengthen their fire code provisions.  Most local jurisdictions in the US and Canada  have \textit{not} adopted the IRC's requirement for sprinklers in small-scale residential construction, but \textit{do} require sprinkler systems for all but the smallest commercial buildings and for certain types of high-density residential buildings, such as highrises. Building codes in China also require sprinkler systems in large or tall commercial buildings \cite{china_sprinkler}. NFPA's and NFPA-13's influence is global, and the discussion that follows is based on the IBC and FPDA 13.

The sprinkler system will use water to put out fires unless the system is in a high-hazard area where another material would be more effective, such as  an aircraft hangar. The water can be in the pipes connected to the sprinkler heads all the time (a \textit{wet} system) or only in case of fire (a \textit{dry} system).  A wet system is the preferred approach unless the pipes are subject to freezing or the area being protected contains valuable material that could be badly damaged by water, such as in a computer room or an art museum. In a dry system, the pipes are pressurized with air or nitrogen; if a sprinkler head activates, the gas rushes out and then water follows behind. 

The IBC dictates sprinkler system requirements according to the use of the area that the sprinkler system will protect. 

Office and residential space are normally  light-hazard, while factories, restaurant kitchens, and enclosed parking garages are normally  ordinary-hazard. 
In light-hazard and ordinary-hazard settings, sprinkler heads will normally be spaced approximately 15 feet apart in each direction in each room.  Further, normally there must be a sprinkler head within 6-7.5 feet of the wall around the perimeter of each room. In general, every room except small closets, plus essentially every other type of interior building space (e.g., hallways, atriums, mezzanines), must satisfy these provisions and have at least one sprinkler.

\subsection{Pipe System Layout and Materials}
Commercial buildings normally  have a large  pipe called a \textit{standpipe} that goes vertically through the building, typically through a stairwell. 

Water for the sprinkler system and for other uses can flow through the same standpipe or \textit{riser} (vertical pipe), but typically the sprinkler system in a commercial building will have its own separate standpipe. That way its water pressure is not diluted from other usage, which is important because high water pressure is key for designing a low-cost sprinkler system. For our purposes, the separation of the two systems means that the sprinkler system pipes will not have noise from running water. 

Each sprinkler system standpipe can service up to 52,000 square feet.  
The pipes that connect directly to sprinkler heads are called \textit{branch lines}, and NFPA 13 requires them to be at least 1 inch in diameter.  Branch lines are tied together on a single floor with (usually larger) pipes called \textit{cross mains}. Larger-yet \textit{feed mains} connect the cross mains to the riser or standpipe on that floor.   To minimize construction costs, generally the smallest diameter pipes are used that will supply the required flow rate (typically a minimum of 7 pounds per square inch) at the sprinkler heads furthest from the riser that serves the sprinkler.
To a first approximation,  the number of sprinkler heads that can be serviced by a pipe of a particular diameter is predetermined, whether the pipe is a branch line, cross main, or feeder main.  For example, NFPA 13 limits 1 inch branch lines to service 10 sprinkler heads, about 150 feet. 

Sprinkler systems are normally installed after structural, heating, ventilation, air conditioning, plumbing, and electrical work is completed.  To dodge the  artifacts left by these tradespeople, branch lines and cross mains normally run \textit{underneath} all these other obstacles, with many twists and turns. 
As a result, in an office environment, the branch line that feeds a sprinkler head is normally located just above the  ceiling.

On a single floor of a building, the sprinkler pipes will be laid out  as a tree, a loop, or a grid.

A tree layout provides exactly one path to each sprinkler head from the riser supplying the sprinkler system on that floor. 

In a loop layout, there are two paths to each sprinkler head. In a grid layout, there are many paths. Every building we have  visited in the course of this research, even those whose hallways form a rectangle,  has had a tree layout, so we leave experiments with other topologies to future work.

Water supply lines for fire sprinkler systems can be made of various forms of steel and iron, copper, or CPVC. 

Black (non-galvanized) steel is the most popular sprinkler pipe material today in commercial buildings, and it is also used for natural gas lines.
 
Today CPVC is allowed in many jurisdictions for light-hazard sprinkler systems. CPVC systems are cheaper to build and install, so their popularity is growing, although we did not see them in any building we visited. 

Older buildings tend to have copper water supply pipes for non-sprinkler use, because copper lasts well, is relatively easy to work with, and used to be quite affordable. In many jurisdictions, PEX is replacing copper, because PEX is very flexible, which makes it quicker and easier to install than copper.  PEX melts at too low a temperature for use in sprinkler systems in commercial buildings, as does PVC, which warps just from carrying hot potable water, and so is largely relegated to waste lines. Since any sprinkler system is safer than none, NFPA 13D does now allow PEX in single-family home fire sprinkler systems.

As low-density residential fire sprinkler system requirements are adopted in more jurisdictions, the pipe attack will become quite practical for those settings. Until then, the attack can be used with residential  water supply lines or gas lines, although unlike fire sprinkler systems, these pipes are not normally present in every room.
 
 \section{Transmitter and Receiver Placement Options}
 \label{sec:placement}

The ubiquity of fire sprinkler systems in commercial buildings throughout the developed world makes them ideal for the pipe attack. 
Sprinkler heads are generally spaced  6-15 feet apart in every room of the building, with pipes adjacent to the sprinkler heads, so an attacker's transmitter can be placed within a few feet of any desired location in the building, as long as the attacker has access to the pipe.
Since  sprinklers do not operate unless there is a fire, the water or air inside sprinkler pipes is stationary, and building codes require the water or air inside the pipes to meet strict pressure standards. Thus these pipes provide a very stable medium for wave propagation,  undisturbed by noise from flow-based environmental changes that other  water and gas  pipes are subject to.

\subsection{ Ceiling Types}

Suspended ceilings are most popular in commercial buildings as they provide repair and remodel flexibility. 
An attacker can install a transmitter or receiver on a branch line in a few minutes: stand on a desk or ladder, push up the acoustic tile next to a sprinkler head, superglue the device to the top of the sprinkler pipe, and drop the tile back in place.  The acoustic tile even provides additional soundproofing, so subsequent data transmission can safely create more ambient noise than with an open ceiling.

An employee, visitor or an attacker disguised as a maintenance worker, tradesperson, or janitor can install the device,  bribe  such a worker to install it, or convince the worker that he is installing a monitoring device for pipe health or air quality.

To maintain an attractive appearance when the ceiling is finished with drywall or another material, normally each sprinkler head is concealed by a color-matched cover plate that will be pushed out of the way by water if the sprinkler head ever activates. The finished ceiling prevents an attacker from easily reaching the branch line to install a transmitter or receiver. 

Instead, the  attacker may be able to reach a fire sprinkler pipe through an existing hole in the finished ceiling that was cut for another mechanical system, such as a modern recessed light that can be popped back up through the ceiling, leaving a hole wide enough to reach through.
If the ceiling offers no convenient holes, the best approach may be to bribe or masquerade as a tradesperson, use a hole saw drill attachment to drill through the drywall near a sprinkler head, reach through the hole to install the device, then pop on a cover plate or a device such as a smoke detector to cover the hole.  

Factories and very modern and very old-fashioned spaces often have an open ceiling, where pipes and wiring are out in the open instead of being concealed behind other materials. With a ladder of the right height, pipe access is easy in these cases.  

\subsection{Receiver Placement Options}

Once a transmitter is in place, the attacker's receiver must be installed somewhere on the same fire sprinkler system. All the sprinkler pipes attached to one standpipe form a single interconnected network, and with a sufficiently strong signal from the transmitter, the attacker should theoretically be able to transmit between any two points in that network. To minimize construction costs,  buildings are designed with as few separate sprinkler systems as possible for their footage. In a building of less than 52,000 square feet under the IBC, all the sprinkler pipes will be on a single system.

In buildings larger than 52,000 square feet, usually construction costs will be minimized by putting a riser in each stairwell of the building, and using the riser to supply water to the sprinkler heads in the 52,000 square feet nearest to that stairwell, with similar coverage on each floor.  This means that an attacker can, in theory, place the receiver in any room relatively close by on the same floor, or in  rooms  above or below it on other floors. 

Our experiments in later sections show that it appears to be practical to place the receiver anywhere on the transmitter's fire sprinkler system {\it on the same floor as the transmitter}.  The risers  in multistory buildings are much larger in diameter than the branch lines, cross feeds, and main feeds that traverse each floor, which makes the risers poor waveguides for the frequencies that we tested.  Further, depending on the frequencies used for transmission, sounds may escape the pipes, echo, and be amplified by the hard surfaces in the stairwell, making them more noticeable to passersby. Thus we do not recommend the use of standpipes for the pipe attack. 

To transmit between floors, non-sprinkler pipes may be more useful. To minimize construction costs, generally water coolers, restrooms, and maintenance closets will be stacked one above the other with narrow-diameter water supply lines running vertically between them. Attacker access to water supply lines is somewhat awkward (e.g., underneath restroom sinks), but
their modest diameter and vertical configuration make the water supply lines a more effective transmission route than fire system risers for the frequencies we tested (1Hz-20KHz).  
However, water supply lines do have the disadvantage of intermittent noise from water flow, and the need for the attacker to install additional equipment that can pass data between the fire sprinkler system and other water supply lines. 

Putting these two options together, when an attacker needs to send signals downstairs, the attacker can transmit signals through the fire sprinkler system to a receiver in a restroom on the same floor, transfer the data from that receiver to a transmitter on the water supply line there, then transmit to a receiver in a restroom on a distant floor.

\subsection{Pipe Interior versus Pipe Surface Mount}

Installing the transmitter or receiver inside the pipe is impractical in terms of access and retrieval.

With an inside-the-pipe mount, attackers also would face the long-term challenge of supplying power to the devices, and sprinkler pipes do not normally contain moving water that might be harnessed for energy. 

The outside mount offers additional potential advantages for battery replacement, device swapping, and data removal, and allows the transmitter and receiver to be removed at the end of the attack.
Given these considerations, we assume that the attacker will mount the transmitter and receiver  on the outside of the pipe. 
\label{Pipe_covert_channel}

\graphicspath{{figures/}}
\section{Getting Data to and from the Transmitter and Receiver}
\label{sec:toandfrom}
In a typical office environment, the transmitter and receiver will be above a dropped ceiling, hidden behind sprinkler pipes.  That raises the question of how to get data to and from the transmitter and receiver.  While this is not the focus of our research, we outline possibilities in this section, and discuss means of detecting them in Section \ref{sec:defenses}. No doubt additional and better options for data transfer to the receiver reside in the arsenal of spy devices used by nation-states. 

The attacker could rely on data sent to the transmitter from small surveillance devices, such as  tiny standalone cameras and audio recorders as well as those incorporated into light bulbs, smoke detectors, clocks, USB plug receptacles, entry door viewers, and pens -- all available on Amazon.
Alternatively, the attacker could personally upload data to the transmitter from the room in question, using a wireless connection.  This requires access or a confederate.
A third option is for the data to be exfiltrated to be already included in the transmitter at the time it is installed.  For example, files could have been loaded onto the transmitter in advance from a computer, e.g., using a USB connection. Alternatively, a camera or audio recorder incorporated into the transmitter could have been used to capture the data before the transmitter was installed.

Retrieving data from a receiver should be easier, as the attacker will have chosen the receiver's location with that need in mind.  For example, the receiver may be near a sprinkler head or sink in a public restroom,  maintenance closet, or in an office that the attacker has rented on the same floor as the transmitter.  If the attacker cannot directly access the receiver through legitimate access to the building or by posing as a tradesperson or maintenance worker, then the attacker will need to hire, bribe, or trick such an individual to change out the receiver or extract its data.

\section{Device Camouflage}
\label{sec:camouflage}

\begin{figure*}[!h]
\centering
\includegraphics[width=2in]{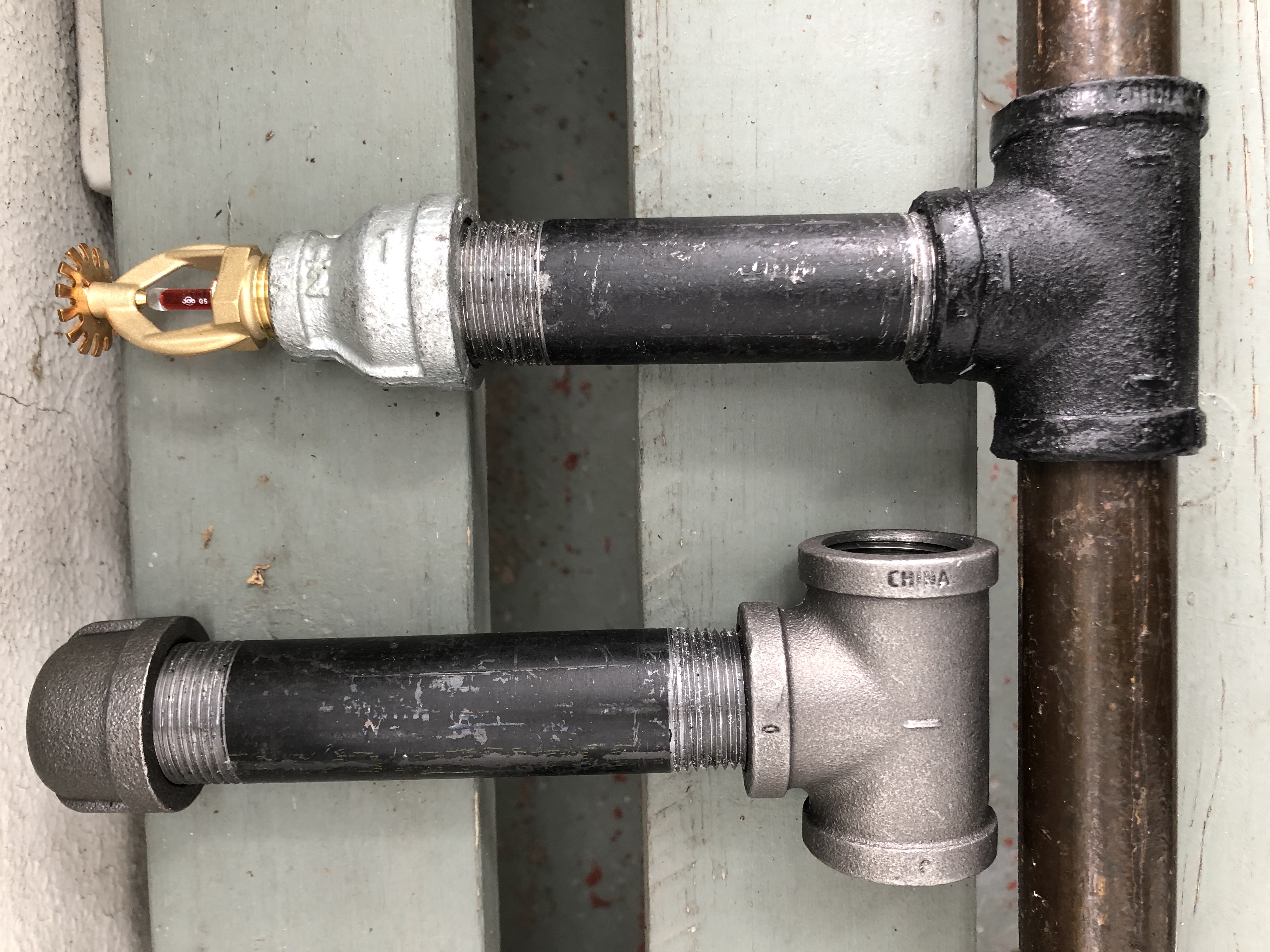}
\includegraphics[width=2in, height=1.5in]{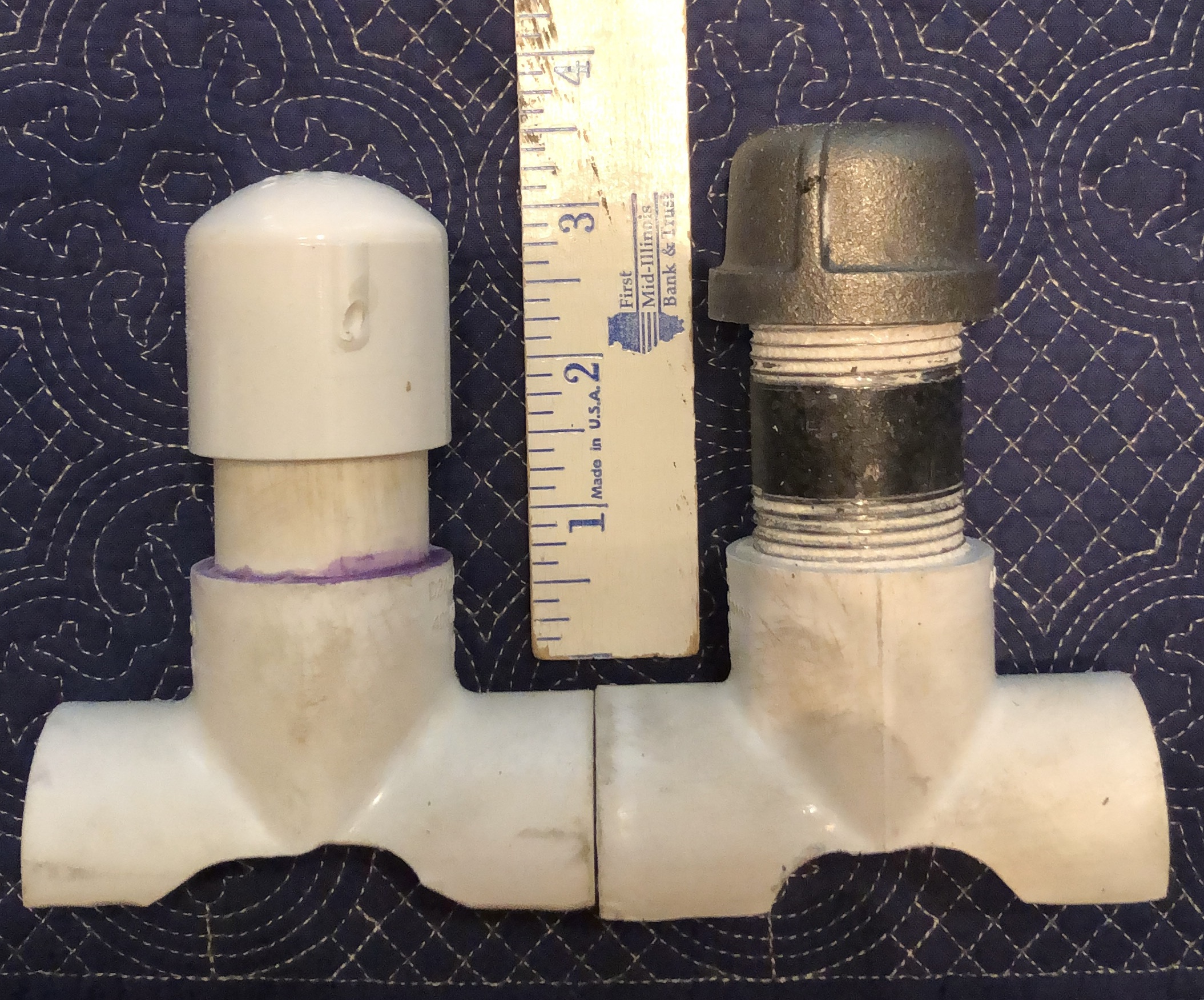}
\includegraphics[width=2in]{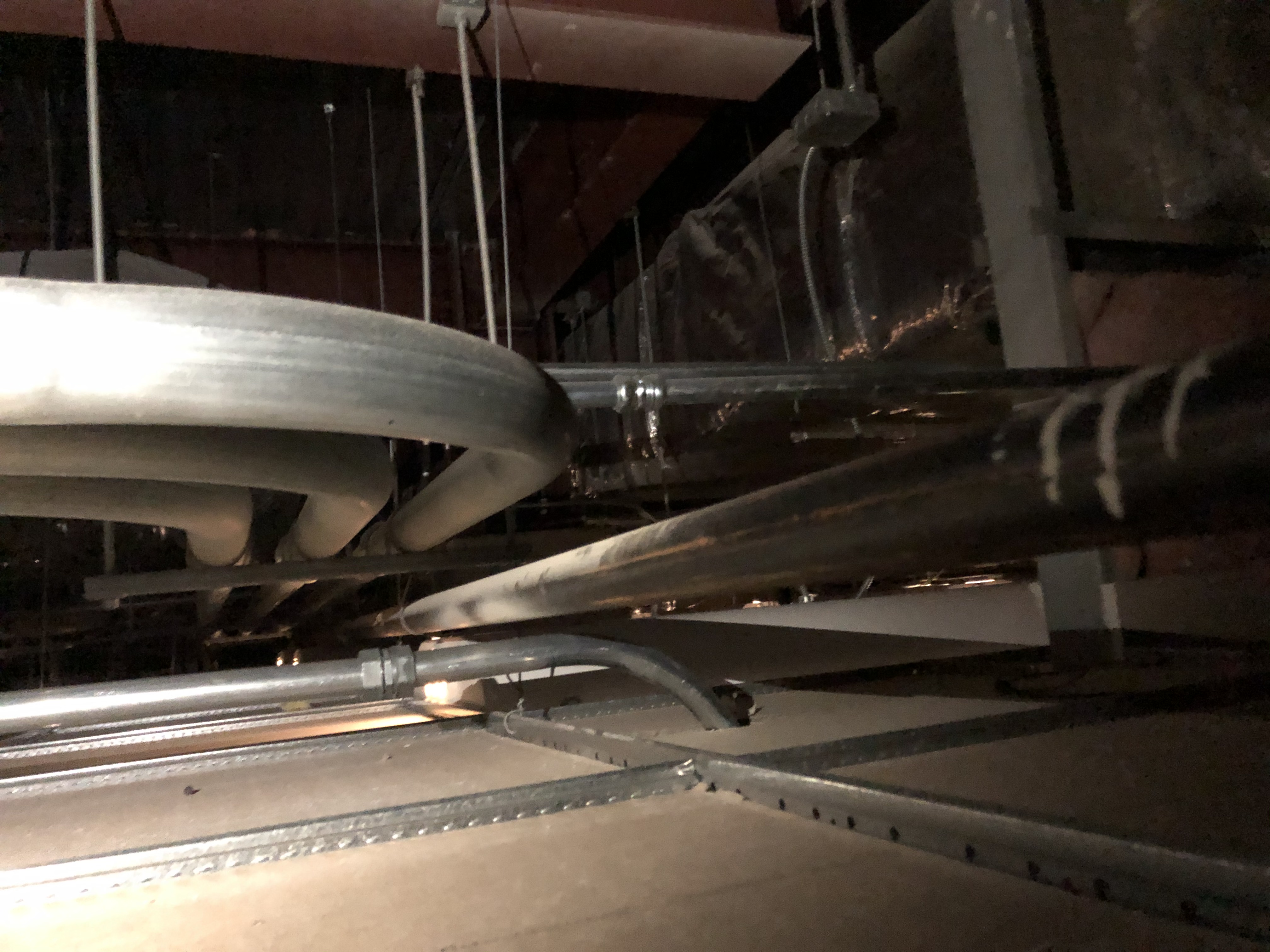}
\includegraphics[width=6in,height=1.5in]{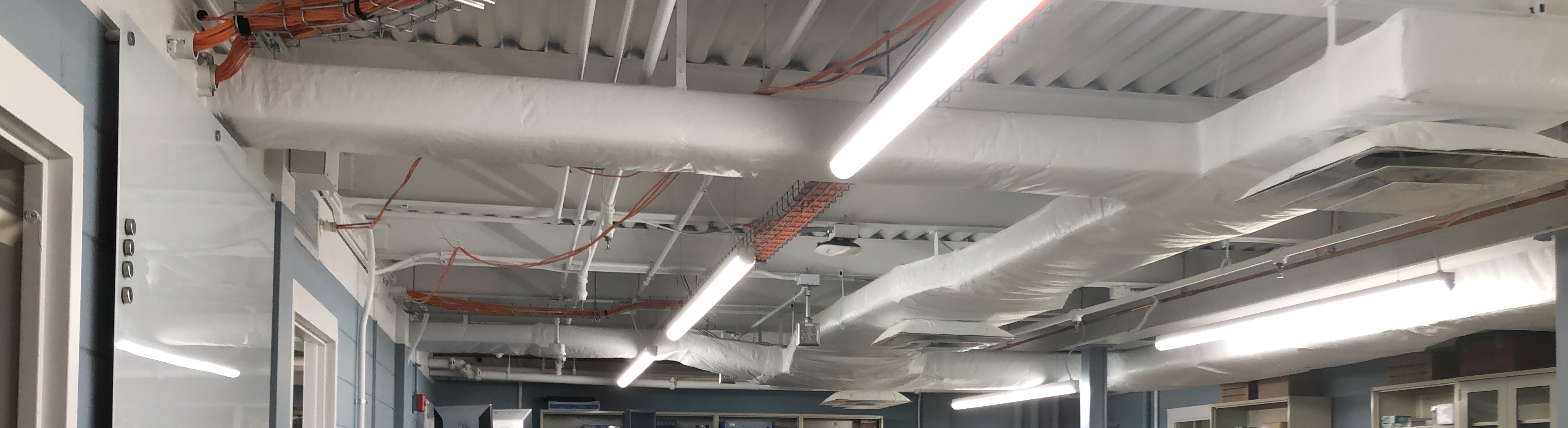}
\includegraphics[width=2in, height=2in]{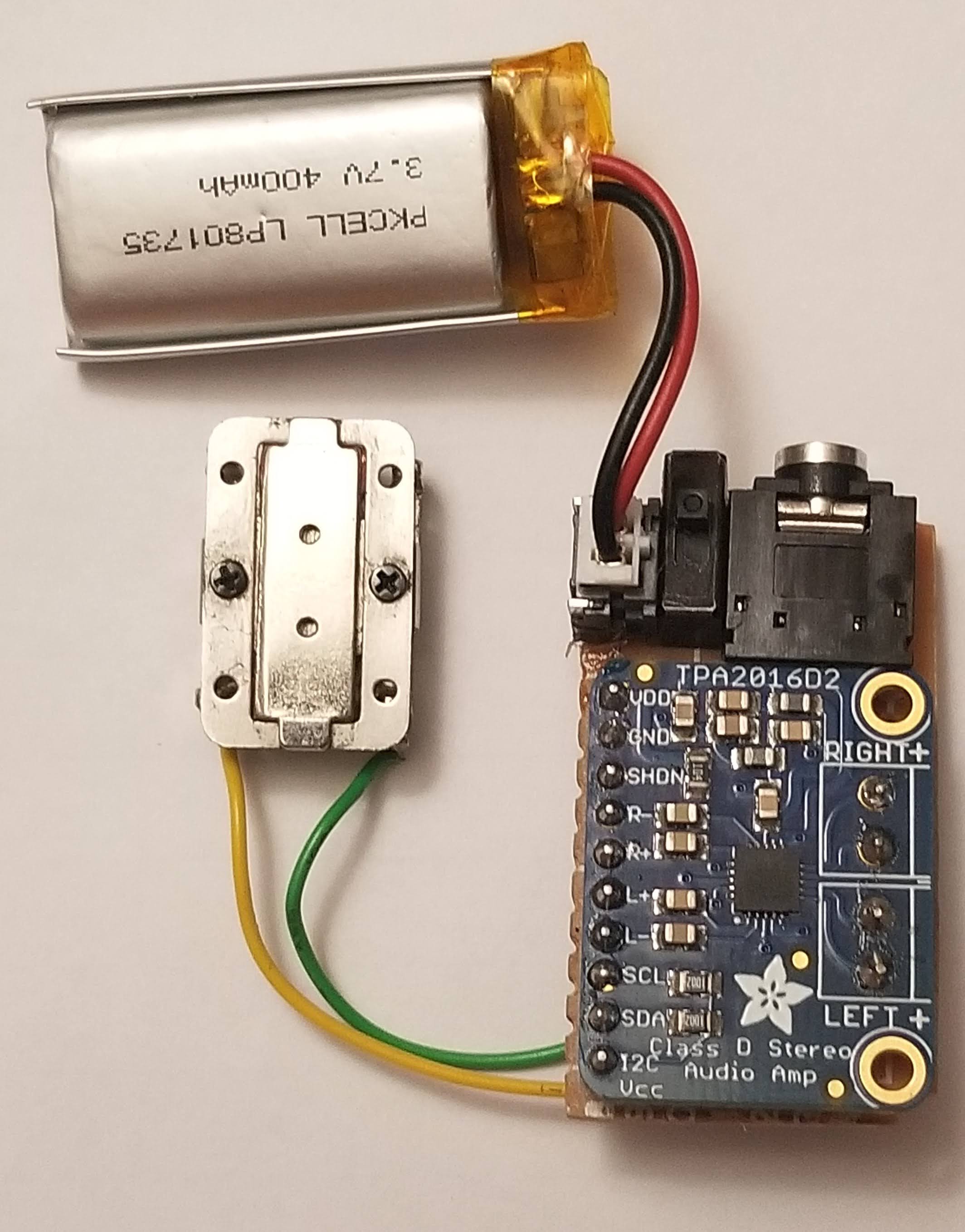}
\includegraphics[width=2in, height=2in]{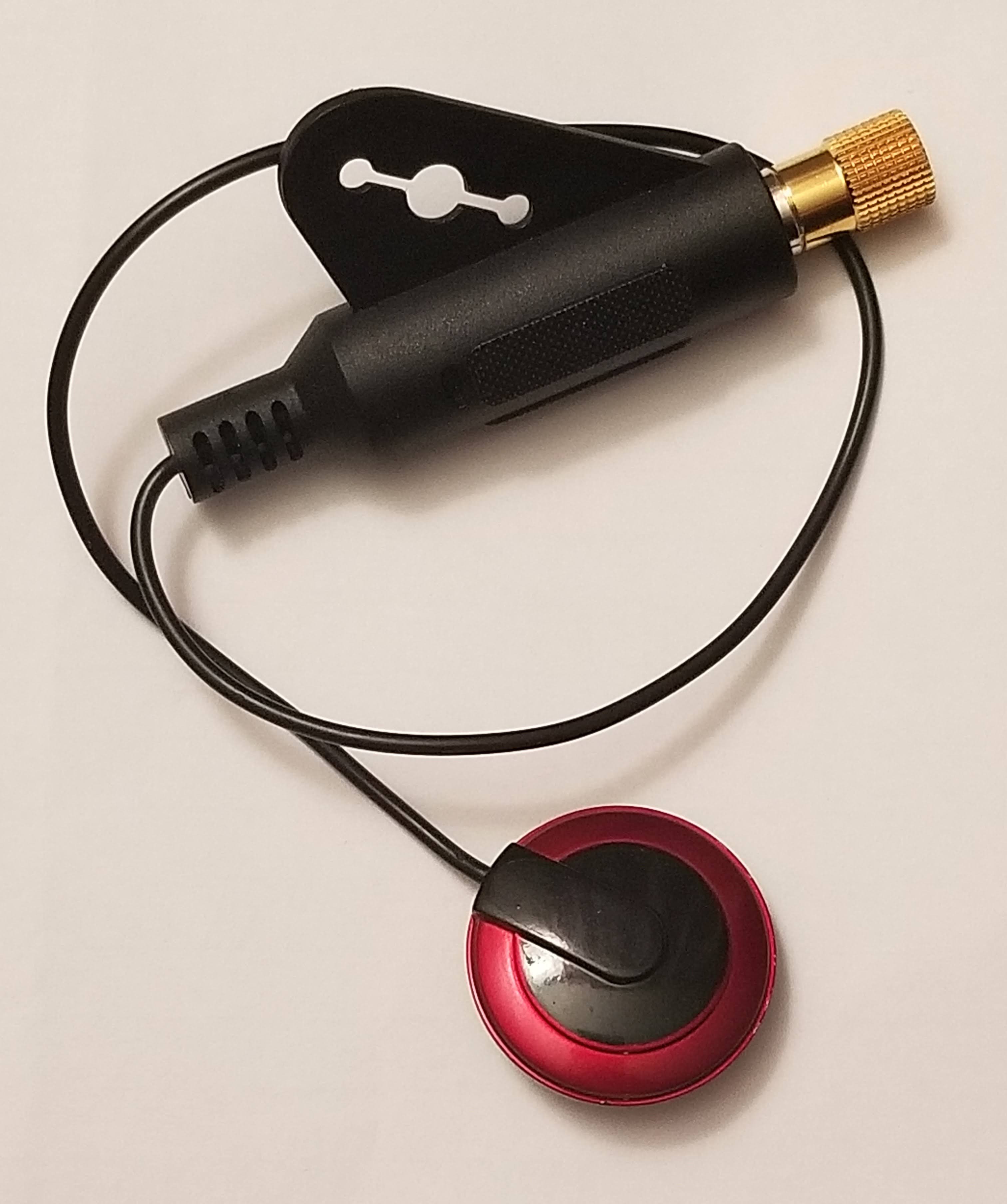}      
\caption{Pipes, Fittings, and Hardware. (top left) In spite of the wide variety of colors, textures, and sheens, all of these are black steel except the sprinkler head itself, the bright silver galvanized steel bell fitting attached to the sprinkler head, and the faux T fitting at upper right. (top center) Pipe length required to hold receiver, mocked up in 1 inch diameter PVC and black steel. (top right) Typical mechanical systems above a suspended ceiling in new construction. (middle) Typical mechanicals below a finished ceiling in old construction. 
(bottom left) Best transmitter tested (Adafruit transducer and amplifier). (bottom right) Best receiver tested (Traderplus).
}
\label{fig:devices}
\label{fig:colors}
\label{fig:pipeLength}
\label{fig:pipeViews}
\label{fig:rockwool}
        \label{final_tx}
        \label{final_rx}
\end{figure*}

We designed the transmitter and receiver to be as small as possible given the supplies and equipment readily available to us at low cost over the internet.  An attacker with more financial resources, patience, and skill would be able to construct much smaller versions of these devices.
But even at their current size,  the transmitter and receiver are easily camouflaged to look like ordinary parts of a sprinkler system.  

As shown at top left in Figure \ref{fig:colors},  pipes and fittings  made of  the standard material for sprinkler system pipes,  black steel, differ greatly in color and sheen, ranging from medium brown through bright gray to black.
We chose one particular fitting to imitate, a  steel T fitting that fits on the same 1-inch diameter pipe used for most branch lines.  We built a removable and repositionable black faux version of the fitting out of an ordinary 1-inch PVC T fitting, as follows. We cut it in half so that it could be attached to and removed from existing pipes. We sanded off the fitting's 3-dimensional markings and visible seams,  added faux 3-dimensional markings and seams identical to those found on black steel T fittings, and added texture and color to make the faux fitting look like a typical black steel T fitting. We created a hinge using the material that forms the thickened bands at the ends of the fitting. The result is shown at top right in the top left photo in Figure \ref{fig:colors}. Only upon touching the fitting does it become apparent that the fitting is not made of black steel.

When installed on the pipe, the hinge seams that allow the fitting to be attached to and removed from a branch line look like ordinary pipe seams.  

In buildings with open ceilings, the building owners often choose to paint the pipes and their hangers (bottom left photo in Figure \ref{fig:colors}), and the faux fitting should be painted to match.

A short length of black steel or camouflaged PVC pipe fits into the camouflaged T, to hold the transmitter or receiver.  As shown at top left and top center in
 Figure \ref{fig:colors}, the pipe can end in a cap, a plug, or  a (non-functional) sprinkler head, pressure gauge, or other device. 
 The top center photo in Figure \ref{fig:pipeLength} shows the shortest length of pipe that held the receiver we built, roughly 3.5 inches, using black steel and uncamouflaged PVC.  Most of the 3.5" of pipe length was needed to hold the batteries for our unit, which were separate from the main circuit board, as shown at bottom center in the figure. 

 An implementation with integrated batteries could use a much shorter pipe than ours did. The top center photo shows a snapfit PVC T fitting as the base, which allowed us to insert and remove the transmitter without removing the cap; however, a real installation must use a faux T fitting instead of a snapfit. 
 
 A trained eye will recognize the oddness of a pipe stub being present at such a random location in the sprinkler system.  To anyone else who pokes their head above the suspended ceiling, even in a newly constructed building (top right photo in Figure \ref{fig:pipeViews}), the faux fitting and pipe will look like just another pipe in a morass of mechanicals. Even with an open ceiling, the faux fitting and pipe stub will fit into the crowd of mechanicals (bottom left photo). 

\section{Hardware Choices}
\label{sec:attackdesign}

\begin{figure*}[!h]
\centering
\includegraphics[width=6in]{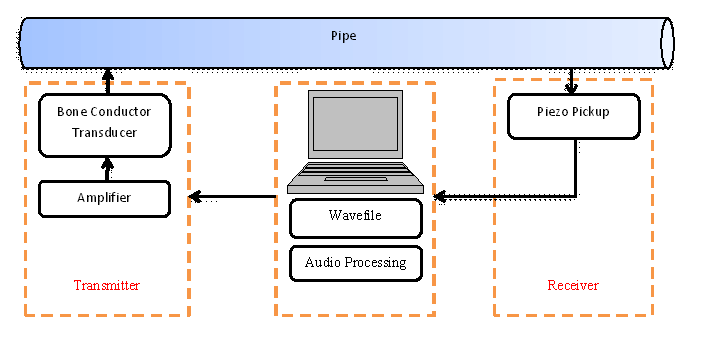}
\caption{Pipe Attack Setup with Adafruit Transmitter (Bone Conductor+Amplifier) and Traderplus Receiver.}
\label{fig:expt_setup4}
\end{figure*}

We conducted experiments to determine what form of signals would be most effective while still affordable for us.
We  determined that acoustic transmissions through mini-speakers leaked much of their sound into the nearby air.

The better option was to send vibrations through the pipe by attaching a transducer that converts audio signals to mechanical vibrations. We found that different models of bone conduction headphones, which are essentially electro-mechanical transducers that produce subtle taps on the surface of the  pipe, had little ambient leakage when used as transmitters. The easily affordable small hardware devices accessible to us send signals up to 20kHz; in a real attack, we would prefer devices that continue on up into the ultrasound range, which would provide additional bandwidth above the range of human hearing.

Given the 20kHz constraint, our final hardware configuration, shown in Figure \ref{fig:devices},  was:

\begin{itemize}
    \item Transmitter: Adafruit 8 ohm bone conductor transducer (14mm x 21.5mm) with TPA2016 Adafruit Class D audio amplifier with AGC (22mm x 28mm)
    \item Receiver: TraderPlus Piezo professional contact microphone pickup, intended for 
    guitars and violins
    \item Processor: Lenovo Thinkpad X230 laptop (Intel i5-3320M CPU @ 2.60GHz)
    \end{itemize}
    
Figure \ref{fig:expt_setup4} shows the logical connection of the Adafruit transmitter and Traderplus receiver to the pipes and the processing unit. The arrows indicate the direction of flow of audio data through the system. The transmitter translates and transmits an audio clip into mechanical taps on the pipe surface. 
The pipe conducts these vibrations to the other end where the receiver circuitry using the piezo pickup is flush mounted on the pipe surface. The pickup receiver is connected to the processing laptop through the audio port. Signals picked up by the receiver are sent to the laptop, which records it and performs postprocessing to analyze the received audio clip. 

We flush mounted the  transducer and receiver on the pipes using a self-adhering bandage. A real attack should instead use superglue for better contact with the pipe and minimal footprint, but superglue proved impractical for us due to the need for frequent repositioning during experiments. 

Since there is a single audio port in the Lenovo laptop, we used a speaker-headphone splitter  at the audio port to split the audio output to the transmitter from the audio input coming in from the receiver. We wrote multithreaded code in Python  to separate the play and record threads in the laptop, synchronized to start the play first followed by the recording in parallel for each test file. We created and analyzed FFTs and spectrograms of recordings, and extracted the frequency and magnitude of the major signal peak in the FFT for each single-tone audioclip.

We also experimented with two Aftershokz bone conductor models. The Aftershokz AS600 wireless bone conductor headphone had a low Bluetooth frame rate, limiting its transmission quality and rate.
The wired Aftershokz Sportz Titanium produced similar audio signal magnitudes  as the Adafruit bone conductor for most frequencies sent along a 3-foot pipe, but with significantly more noise than the Adafruit. 


We also experimented with potential  receivers from nine other manufacturers. Six of these devices were intended for music applications, one for medical, and one for mechanics. Some of the other receivers were less sensitive, some picked up more environmental noise, some added more noise to the circuit themselves and some broke so easily that they were not suitable for use in experiments that required repositioning of the device in different buildings and at different distances along a sprinkler system.

We compared  a laptop and a Samsung Galaxy phone as the processing unit, and found that the laptop injected less noise into the received audio.

\section{ Buildings and Pipelines Tested}
\label{sec:buildings}

Our experiments used six pipe runs in three buildings: an old commercial building with both wet and dry sprinkler systems (Old Commercial), a new commercial office and laboratory building with wet sprinklers (New Commercial), and an old single-family residence with new plumbing (Residence). Table \ref{table:attack_locations} summarizes the test locations, including pipe types, sizes, and lengths.  

\begin{figure}[!htb]
\centering
\includegraphics[width=3in]{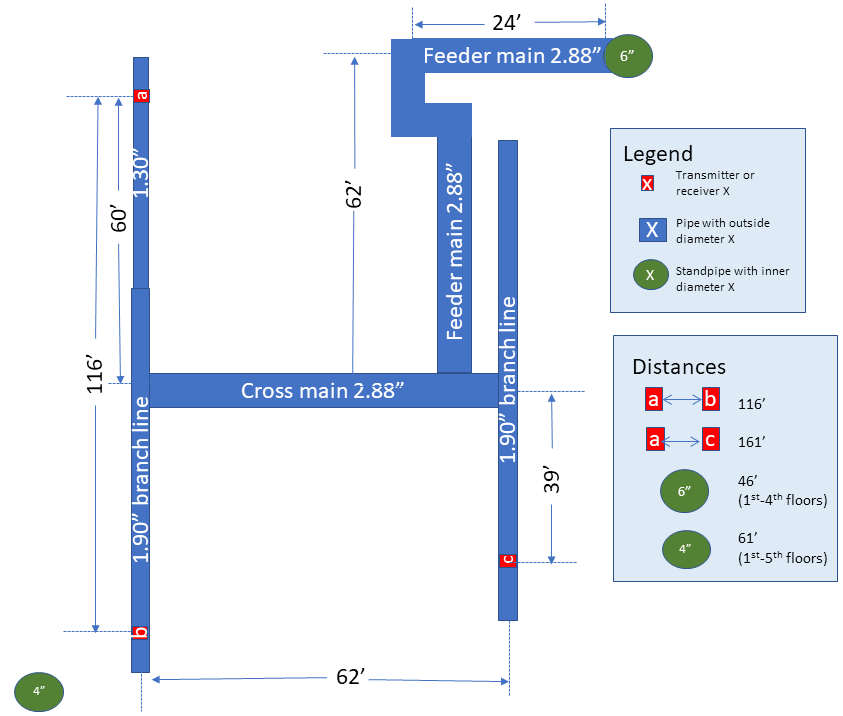}
\caption{Layout and Outer Diameter of Fire Sprinkler Pipes Tested at New Commercial Building Locations NCS and NCR. Not shown are the dozens of branch lines that intersect the pipe pathways used in the tests, or the many small twists and turns to dodge other mechanicals.}
\label{fig:siebel_layout}
\end{figure}

\textbf{New Commercial.}  Built in the 2000s, this  commercial building has over 200,000 square feet of space spread across four floors and a basement, as well as a partial fifth floor. As befits its size, the building has multiple separate sprinkler systems, each made of black steel and each laid out in a tree topology. We conducted our experiments in one wing of the building, which is served by a single sprinkler system that we estimate as covering 50,000 square feet.  This wing primarily has dropped ceilings with acoustic tiles, and houses office and lab space with a relatively open floor plan.  We mapped the topography of the sprinkler system in the wing, which follows an almost identical pattern on all floors. We took measurements on one floor of the wing and in two of the stairwells near the wings, one of which housed the riser that served the wing, and the other of which held a smaller riser that served another section of the building. Figure \ref{fig:siebel_layout} shows the relevant portion of this building's sprinkler system layout.

\textbf{Old Commercial}. Built in the 1910s, this commercial building has been remodeled and repurposed multiple times over the years.  Today it offers a modern open floor plan with open ceilings and a 40,000 square footprint on each of two floors and a basement. The basement and first floor have a wet sprinkler system, and the second floor has a dry sprinkler system filled with air. The sprinkler system pipes are black steel. Each floor has a tree topology. 
We took measurements in the basement and on the second floor.  

\textbf{Residential}. Built in the first decade of the 1900s, this residential building has a 2,000 square foot footprint on each of three floors and a basement. The building has no sprinkler system. Water supply lines in the house are copper pipes installed in the 1990s. The natural gas supply lines running along the ceiling of its basement are made of black steel.

 Table \ref{table:attack_locations} includes our empirical measurements of pipe external diameters. The 1.66" outer diameter pipes correspond to a nominal 1.25" inner diameter.  Similarly, 1" internal diameter corresponds to 1.32" outer diameter, and 2.5" inner diameter corresponds to 2.875" outer diameter.

\begin{table}[!htb]
\centering
\begin{adjustbox}{width=\columnwidth,center}
\begin{tabular}{|l|l|l|l|l|}
\hline
\cellcolor[HTML]{D9D9D9}Name &
  \cellcolor[HTML]{D9D9D9}Pipe Description &
  \cellcolor[HTML]{D9D9D9}\begin{tabular}[c]{@{}l@{}} Material \& OD\end{tabular}
  &
  \cellcolor[HTML]{D9D9D9}Pipe Contents &
  \cellcolor[HTML]{D9D9D9}Distances Measured \\ \hline
RG &
  \begin{tabular}[c]{@{}l@{}}Residential Gas Pipe \\ 50ft long, with branches\end{tabular} &
  \begin{tabular}[c]{@{}l@{}}Black Steel\\  1.04”\end{tabular} &
  Natural Gas &
  20ft, 30ft, 40ft \\ \hline
RW &
  \begin{tabular}[c]{@{}l@{}}Residential Water Supply \\ 40ft long, with branches\end{tabular} &
  \begin{tabular}[c]{@{}l@{}}Copper\\    1.04”\end{tabular} &
  Water &
  20ft, 30ft, 40ft \\ \hline
OCS &
  \begin{tabular}[c]{@{}l@{}}Old Commercial\\  Wet Sprinkler \\    170ft, with branches\end{tabular} &
  \begin{tabular}[c]{@{}l@{}}Black Steel\\    1.05” to 1.65”\end{tabular} &
  \begin{tabular}[c]{@{}l@{}}Water \\ (pressurized)\end{tabular} &
  \begin{tabular}[c]{@{}l@{}}20ft, 40ft, 60ft, 80ft, \\ 100ft, 120ft, 170ft\end{tabular} \\ \hline
OCDS &
  \begin{tabular}[c]{@{}l@{}}Old Commercial\\  Dry  Sprinkler \\    90ft, no branches   
  \end{tabular} &
  \begin{tabular}[c]{@{}l@{}}Painted Steel Pipe \\ 1.93” \end{tabular} &
  Air &
  20ft, 50ft, 60ft, 90ft \\ \hline
NCS &
  \begin{tabular}[c]{@{}l@{}}New Commercial \\ Wet Sprinkler  \\ Multibranch with  crossmains\end{tabular} &
  \begin{tabular}[c]{@{}l@{}}Black Steel \\    1.30” to 1.90” \\ branch lines,    \\ 2.88”  crossmain\end{tabular} &
  \begin{tabular}[c]{@{}l@{}}Water\\  (pressurized)\end{tabular} &
  \begin{tabular}[c]{@{}l@{}}10ft,  50ft,  80ft, 116ft \\ (adjacent branch lines) \\   \\ 161ft \\ (non-adjacent branches \\
  separated by crossmain)\end{tabular} \\ \hline
NCR &
  \begin{tabular}[c]{@{}l@{}}New Commercial \\  Riser (wet) \\  6-floor standpipe\end{tabular} &
  \begin{tabular}[c]{@{}l@{}}Painted Steel \\    4”\end{tabular} &
  \begin{tabular}[c]{@{}l@{}}Water\\  (pressurized)\end{tabular} &
  30.3ft,  45.6ft,  61.2ft \\ \hline
\end{tabular}
\end{adjustbox}
\caption{Measurement Locations for the Pipe Attack}
\label{table:attack_locations}

\end{table}

When appropriate buildings were not available to us for a particular experiment, we built our own pipe run. For example,
CPVC is approved for use in some jurisdictions for fire sprinkler systems, but no building we had access to used CPVC pipes for any purpose, so we built our own CPVC pipe run.  

More precisely, we built  two indoor pipelines, each approximately 30 feet in length. One used 3/4" inner diameter CPVC (1" diameter was not locally available) and the other of 1" PVC. We outfitted these with a variety of PVC, metal, and CPVC fittings and filled them with unpressurized air or water. We used solvent-based connections wherever possible.

We also built an outdoor pipeline roughly 42.5 feet long, using two 21 foot lengths of 1.25" inner diameter fire-rated Schedule 10 black steel pipe (both 1" and 1.25"  are common in branch lines).
These pipes have a groove joint at the ends, which we used to connect them to several models of fire-rated Victaulic  couplings. We also used grooved couplings to connect the pipes to the fittings and adaptors required to connect to a water source and valve/drain.

\label{Experimental_setup}

\graphicspath{{figures/}}

\section{Experimental Results}
\label{sec:experimentResults}

We evaluated several characteristics of the communication between the transmitter and receiver:
\begin{itemize}
    \item Best frequency bands for each pipe run, and bandwidth available
    \item Attenuation and/or signal strength changes for different pipe runs over distances
    \item Covertness of setup and environmental signal leaks
\end{itemize}

\subsection{Frequency Response and 
Bandwidth}
In theory, the attacker can broadcast information in parallel on all frequencies   within the range of the  transmitter and receiver.
However, some frequencies may tend to match environmental noise or be absorbed or distorted by the pipe along the way, leading to issues with attenuation, interference, or even (in audible frequencies) growing loud enough along the way to be noticeable.  

To characterize the ability of a pipe run to act as a communication channel for a given frequency, we evaluated its frequency response, which  determines the bandwidth that the communication channel can support at a particular volume. 
In response to a stimulus signal, frequency response measures the output produced by the system in terms of the amplitude and phase of the received frequencies.
We generated and transmitted lossless single-frequency audio wave files  of 1 second signals, preceded and followed by 0.5 second silence, for  frequencies from 1Hz to 20kHz, at 100Hz intervals, with a sample rate of 44100Hz.

This range is supported by most commercially available audio devices (speakers, headphones, microphones) built for supporting the human audible frequency spectrum, 20Hz to 20KHz. Had higher frequency devices been available to us, we would have extended our investigation into the ultrasound range as well.

We determine whether attackers should use a particular single frequency using the following \emph{Bandwidth Condition}: given pipe noise and other attenuation, when the transmitter sends frequency $f$, is $f$  the major frequency in the signal seen by the receiver? More precisely, is frequency $f$ within 5Hz of being the main peak in the frequency domain, after we perform a Fast Fourier Transform on the received audio clip, at all measurement locations along the pipe?

If so, we include $f$ in the attacker's bandwidth, and determine how much $f$ attenuates or changes in magnitude as it passes along the pipe. 
This is a poor-man's approach, as more complex signal processing methods should be able to extract a wider range of signals that may not arrive as a peak. 
\begin{figure*}[!h]
\centering
        \includegraphics[width=0.45\linewidth,keepaspectratio]{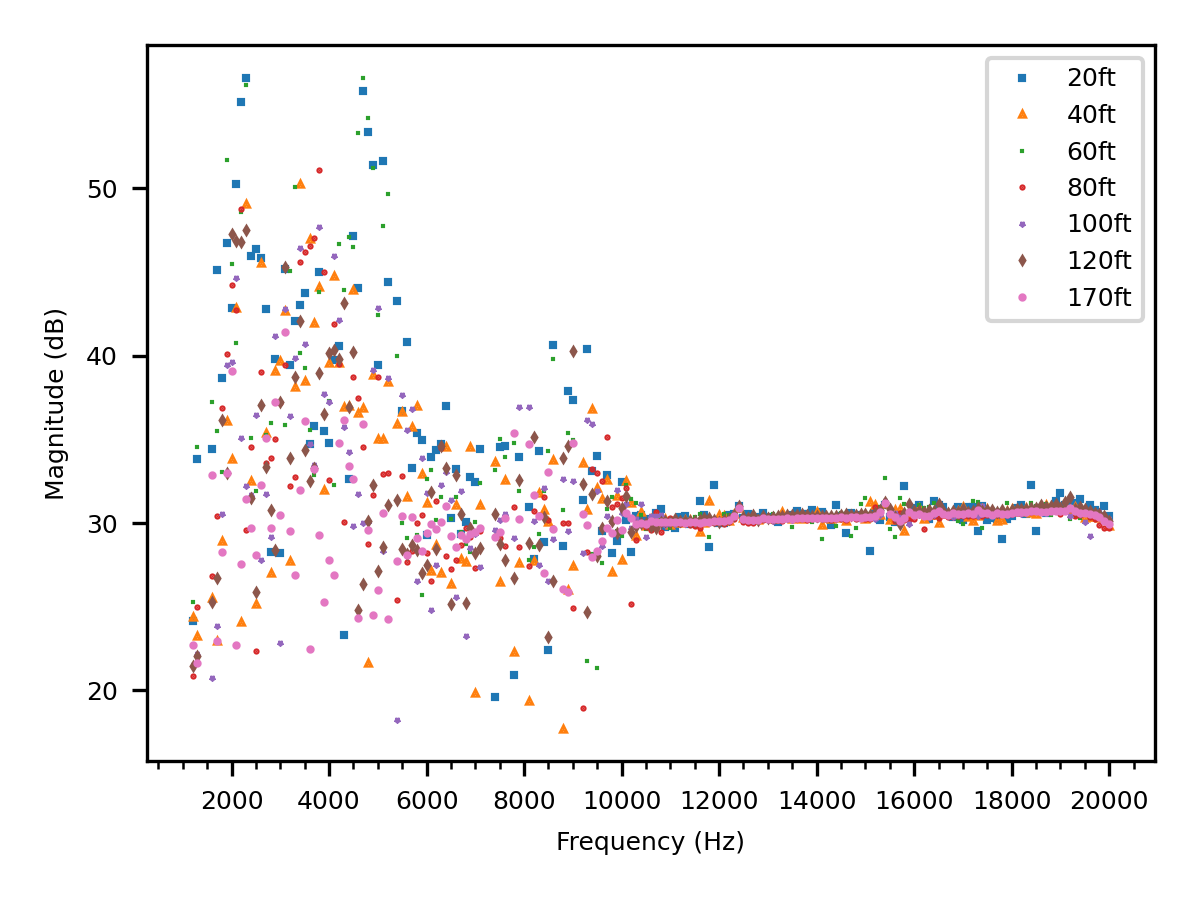}
        \includegraphics[width=0.45\linewidth,keepaspectratio]{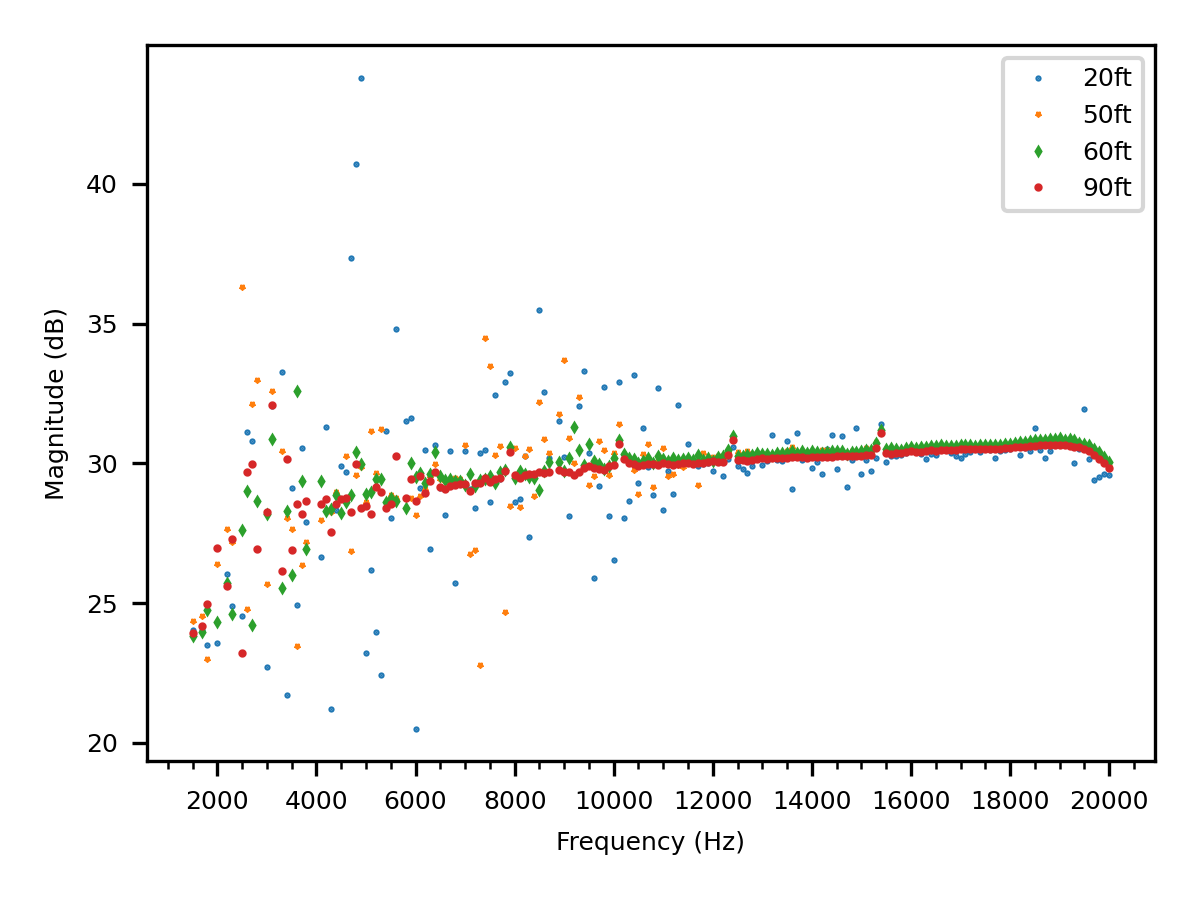}
        \includegraphics[width=0.45\linewidth,keepaspectratio]{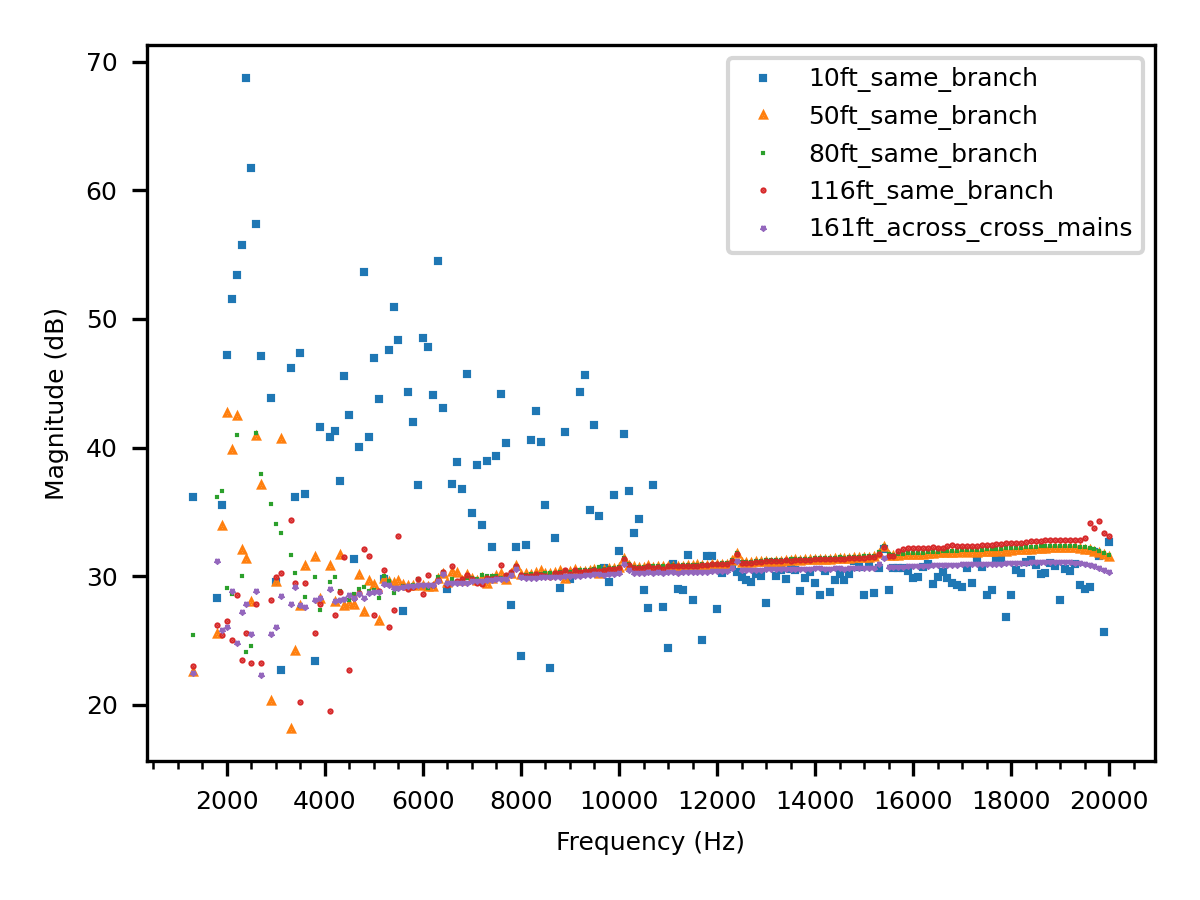}
        \includegraphics[width=0.45\linewidth,keepaspectratio]{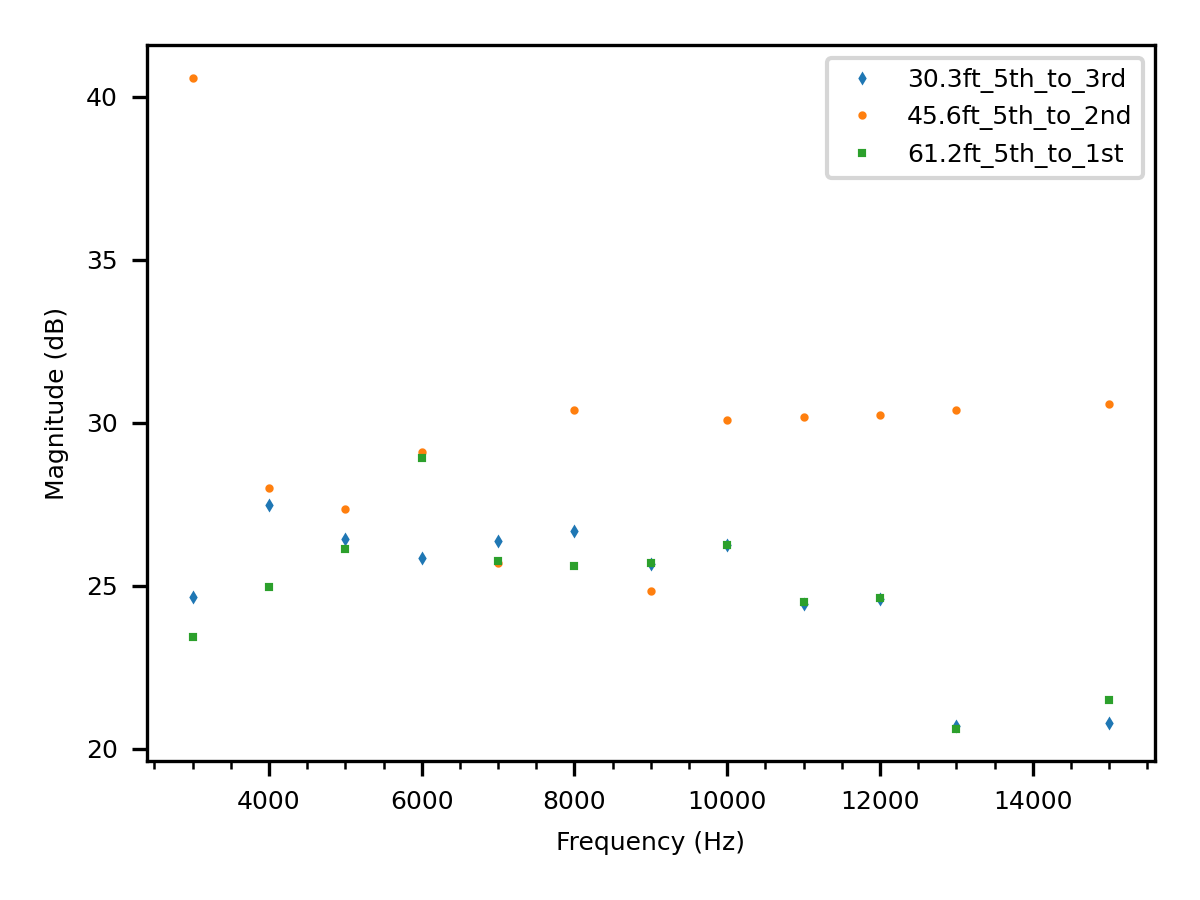}
        \includegraphics[width=0.45\linewidth,keepaspectratio]{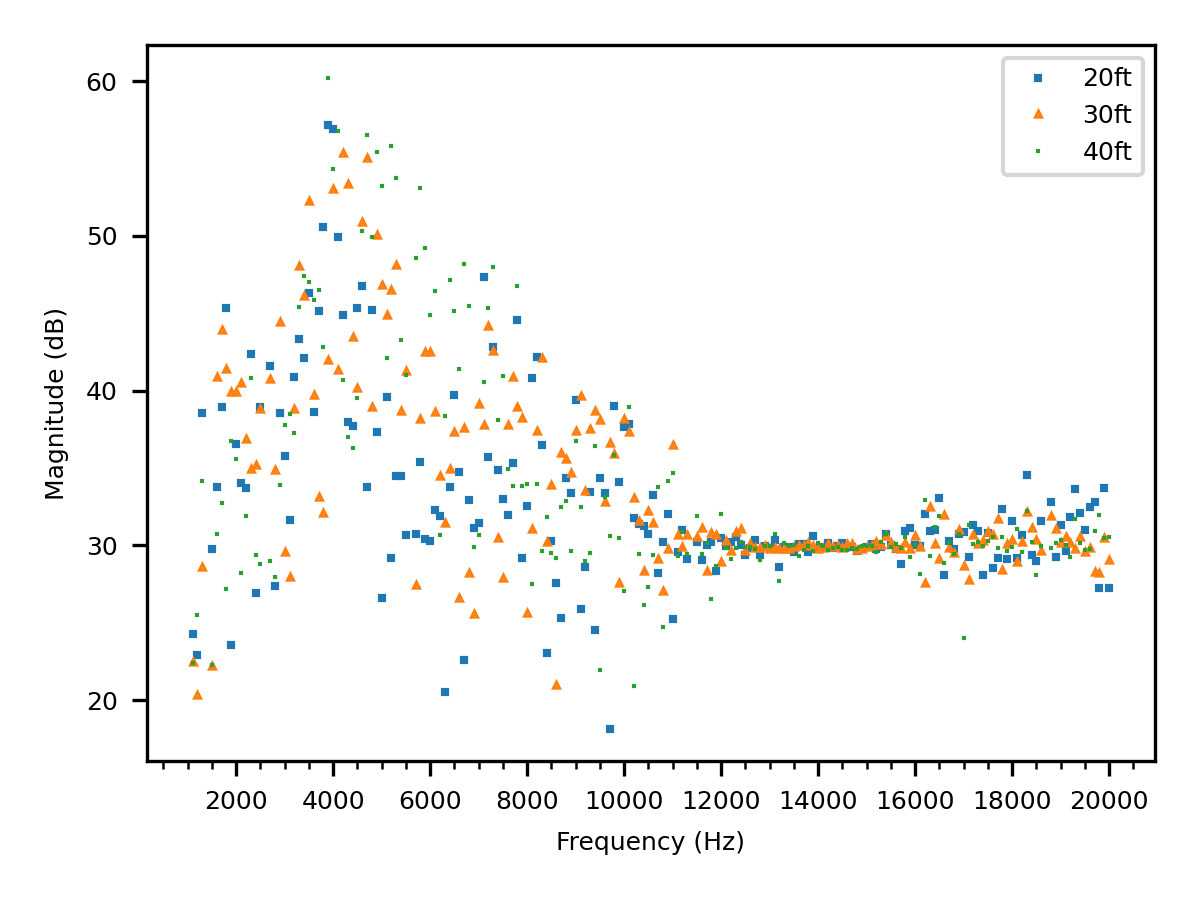}
        \includegraphics[width=0.45\linewidth,keepaspectratio]{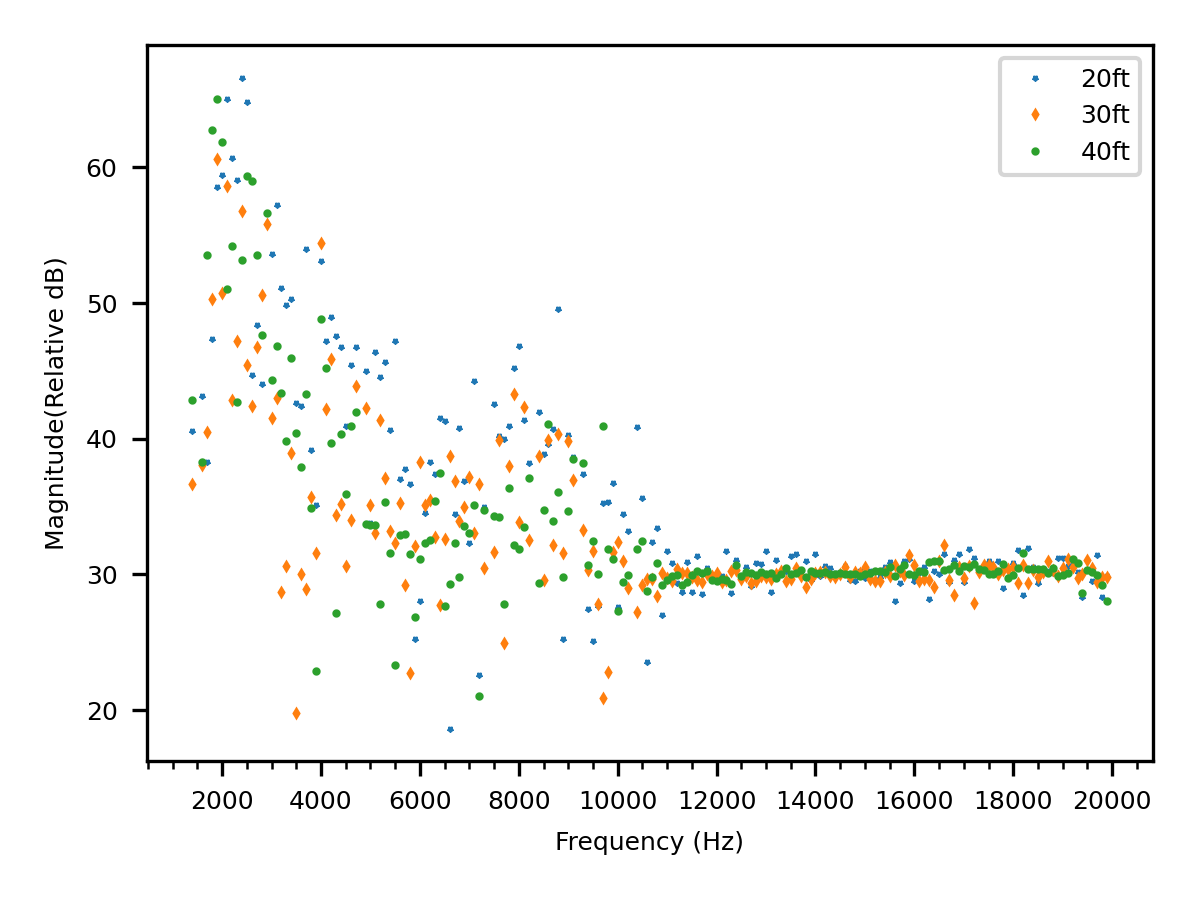}

     \caption{Frequencies Satisfying the Bandwidth Condition After Transmission Through Various Pipe Runs.  
     (top left) Old commercial building wet sprinkler. 
     (top right) Old commercial building dry sprinkler. 
     (middle left) New commercial building black steel sprinkler. 
     (middle right) New commercial building 4" wet standpipe/riser from 1st to 5th floors, every 1000Hz only.
     (bottom left) Residential black steel gas pipe. 
     (bottom right) Residential copper water pipe.}
    \label{fig:building_attacks_newcommercial}
    \label{fig:siebel_standpipe}
    \label{fig:siebel}
    \label{fig:laura_dry}
    \label{fig:laura_wet}
    \label{fig:res_gas}
    \label{fig:building_attacks_oldcommercial}
    \label{fig:building_attacks_residential}   
 \end{figure*}

For the pipe measurement locations described  in Table \ref{table:attack_locations}, Figure~\ref{fig:building_attacks_oldcommercial} shows  all received frequencies that satisfy the Bandwidth Condition, along with their received magnitude.

Though we expected a clear attenuation trend with distance across the pipes, Figure \ref{fig:building_attacks_newcommercial} shows that that is not the case; many frequencies' signals decrease in strength at first but then increase in magnitude as they pass through the pipe. 
We hypothesize that the increases are due to constructive interference between the waves at certain points in the pipe.

In all buildings and pipe runs except the 6" standpipe, we successfully transmitted and received many frequencies at the longest distances available.

The longest pipe run was 170ft in OCS along a black steel wet sprinkler pipe; the pipe run was exposed and was relatively straight, though with many fittings. The second longest but much more twisting pipe run was in the NCS concealed wet sprinkler system, where we  transmitted  from one branch line end,  through a larger cross main, then to the end of the furthest branch off the cross main. Thus the signal transmitted well across  complex pipe structures on a single floor. Further, the de facto limit of about 150' per typical branch line (10 sprinklers of standard coverage, spaced 15' apart) means that an attacker should be able to transmit and receive along the full length of any branch line, as well as anywhere within a single-floor sprinkler system run in a building with four (as in the measured wing of NCS) or more floors. Finally, since the figure shows that signals at many frequencies hold their magnitude well at the maximum distances measured, the maximum potential distance for successful transmission is certainly not limited to 161'.

The least successful transmission in Figure \ref{fig:building_attacks_newcommercial} was for the NCR 4" wet risers, which had a lot of low frequency noise coupled with serious attenuation and ambient leakage issues for the frequencies we tested; only 3KHz-13KHz and 15KHz satisfied the bandwidth condition. The NC 6" standpipe was even worse, and is not shown in the figure. As attackers, we would avoid these pipes. Note also that transmissions on the adjacent floor branch lines and cross main could not be picked up on the risers in the stairwell.

\subsection{Best Frequencies for Attackers}

An  attack should be easy to configure, hard to detect, and have a low signal loss and error rate. This means that a preconfigured attack should use frequencies that are known to work well in most pipes, as details regarding the sprinkler system pipes and layout are usually not available.
The attacker should use frequencies that people are unlikely to overhear, which is trivial with ultrasound but requires attention with equipment like ours.
Finally,  to maximize battery life and reduce the risk of detection, the attacker should use the lowest volume/power that allows successful reception of the signal. 
In theory, the first and last of these three parameters can be calibrated by firmware during attack setup. Here we provide a poor-man's guide to the best choices.

{\bf Frequencies That Work for Most Pipes.}

\begin{figure}
    \centering
    \includegraphics[width=\columnwidth]{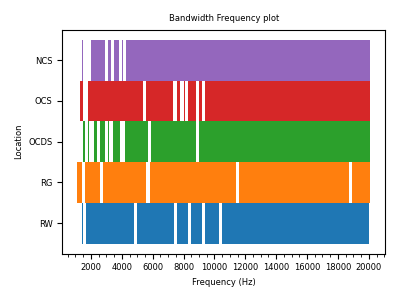}
    \caption{Frequencies That Satisfy the Bandwidth Condition at All Distances. }
    \label{fig:global_attackfreq}
\end{figure}

Some frequencies consistently performed well regardless of pipe material, pipe diameter, pipe layout, building structure, environmental noise in the pipe, and transmission distance. Figure \ref{fig:global_attackfreq} provides this global view of the best bandwidth, i.e., frequencies that satisfied the {Bandwidth Condition} for all  single-floor transmissions in all buildings tested, as detailed in Figure \ref{fig:building_attacks_newcommercial}.

Figure \ref{fig:global_attackfreq} shows that the longest stretch of contiguous good frequencies for all buildings is in the 12KHz to 18KHz bandwidth range (12kHz-20kHz for commercial spaces).

{\bf Frequencies Least Noticeable to Human Ears.}

The human audible frequency range is often considered to be 20Hz-20kHz \cite{humanrange}, with human ears  most sensitive to  2kHz-5kHz \cite{sensitivefrequencies}.  Human ears tend to grow less sensitive as the frequency increases, and the  upper limit of human hearing is typically 15-17kHz \cite{reducedhumanrange, upperhearinglimits}, hence the appeal of ultrasound and near-ultrasound for a stealthy attack. 
We tested this limit with 8 volunteers aged 20s to 60s, and found that

the inaudible frequencies began at  12kHz for 2 volunteers, 14KHz for 1 volunteers,  15KHz for 4 volunteers and 16.3KHz for 1 volunteer. 

Thus an attacker will strategically prefer to use these upper frequency ranges (15KHz+), which Figure \ref{fig:global_attackfreq} shows are also transmission-friendly.

{\bf Lowest Possible Transmission Volume and Power.}

During the initial setup phase of an attack, the attacker could also determine the lowest possible transmission volume that provides good reception, for each frequency that the attacker plans to use. As mentioned earlier, this check could also help to avoid any locally noisy frequencies. For buildings with consistent pipe noise, the attacker could also make an initial recording of the noise and employ noise removal techniques to improve signal detection. However, none of our tested pipes were very noisy at high frequencies, even when a group of children were building with legos in OCS and we expected to have to rerun the experiments.

To estimate the lowest volume/power at which transmissions could be made and still be intelligible above the background noise of a pipe, we transmitted a linear chirp descending from 20kHz to 1Hz along the maximum distance of each building pipe run. For RW and RG, which are 40' residential runs, a minimum laptop volume setting of 6 (peak loudness of 22.8dBA, a bit softer than a whisper) sufficed for the graph of the signal to be clearly visually differentiated from the background noise. For OCS, OCDS, and NCS, a minimum laptop volume of 10 (peak loudness of 30.5dBA, about as loud as a quiet rural area) was required, while the New Commercial 4" standpipe required a laptop volume of 30 or higher.

To  minimize leakage of the  signal into the air at the point of transmission, soundproofing would be important in a real attack, especially if the contact between a flat transducer and the curved pipe is not perfect. Our implementation of the transmitter did not include any soundproofing. In our experience, some signal also leaks into the air along the pipe run, but  the transmitter was the main point of leakage.

\label{Attack_eval}
\graphicspath{{figures/}}

\section{Channel Capacity Analysis}
\label{sec:shannon}
Given the empirical results summarized in Figure \ref{fig:building_attacks_newcommercial}, we can derive an upper bound on the channel capacity for each individual frequency in each pipe run, and for the channel capacity of each pipe run as a whole. To closely approach these upper bounds in practice would require the use of a communication protocol with excellent error correction.

For individual single-frequency channels in the presence of noise, we have the Shannon-Hartley Theorem \cite{taub86}:

\begin{equation}
    C=B \log_2((N+ S)/N),
\end{equation}

\noindent
where $C$ is the channel capacity in bits per second; $B$ is the bandwidth of the channel in Hz; $S$ is the average received signal power over the bandwidth, measured in watts; and $N$ is the average noise or interference power over the bandwidth, measured in watts. 
  
In this equation, we set $S+N$ to be the power spectral density of the entire signal coming through the pipe (i.e., the intended signal $S$ plus the noise $N$). We set $N$ to be the power spectral density of the signal coming through the pipe when no transmission was taking place.  For this we used a .5 second recording of the sound on the pipe before each frequency was transmitted.  Finally, $B$ is the bandwidth of the individual channel being used, e.g., 19kHz.

We focused on the maximum distance transmission for each pipe tested and determined the highest-capacity frequency across all frequencies tested in  Figure \ref{fig:building_attacks_newcommercial}. In all pipes tested except the 4" standpipe, the highest-capacity frequency was close to 20KHz and its capacity was roughly 300K bits/second (36K bytes/second). Table \ref{table:individual_channel_capacities} gives the details.
  
  \begin{table}[!h]
\centering
\begin{adjustbox}{width=\columnwidth,center}
\begin{tabular}{|l|l|l|l|l|}
\hline
\cellcolor[HTML]{D9D9D9}Pipe &
  \cellcolor[HTML]{D9D9D9}Pipe  &
  \cellcolor[HTML]{D9D9D9}Frequency &
  \cellcolor[HTML]{D9D9D9} Capacity &
  \cellcolor[HTML]{D9D9D9} Total\\
  \cellcolor[HTML]{D9D9D9} Run&
  \cellcolor[HTML]{D9D9D9} Type&
  \cellcolor[HTML]{D9D9D9} (Hz)&
  \cellcolor[HTML]{D9D9D9} (bps) &
  \cellcolor[HTML]{D9D9D9} Capacity(Mbps) \\\hline
OCS &
  170' Wet Sprinkler & 19,900 & 307,980 & 13.03

 \\ \hline
OCDS &
90' Dry  Sprinkler & 19,900  & 307,061 & 13.00 \\ \hline
NCS &
  161' Wet Sprinkler & 19,900  & 310,443 & 13.10 \\ \hline
NCR &
 62' 4" Riser & 15,000  & 182,487 & 13.60 \\ \hline
 RG &
   50' Gas Pipe & 19,800  & 313,611 & 12.70 \\ \hline
RW &
  40' Water Supply & 19,600  & 300,718 & 12.80  \\ \hline
  Indoor Lab &
  30' Water CPVC & 19,900  & 303,950 & 12.91 \\ \hline
  Indoor Lab &
  30' Air CPVC & 19,900  & 303,532 & 13.05 \\ \hline
  Outdoor Lab &
  42.5' Victaulic 75 & 19,900  & 380,532 & 16.82 \\ \hline
\end{tabular}
\end{adjustbox}
\caption{For  Each Building Pipe Run, the Highest Capacity Sub-channel Frequency at Maximum Distance, and the Total Channel Capacity (bps)  with 100Hz Frequency Separation in 15KHz-20KHz}
\label{table:individual_channel_capacities}
\end{table}

\begin{figure*}[!h]
\includegraphics[width=55mm]{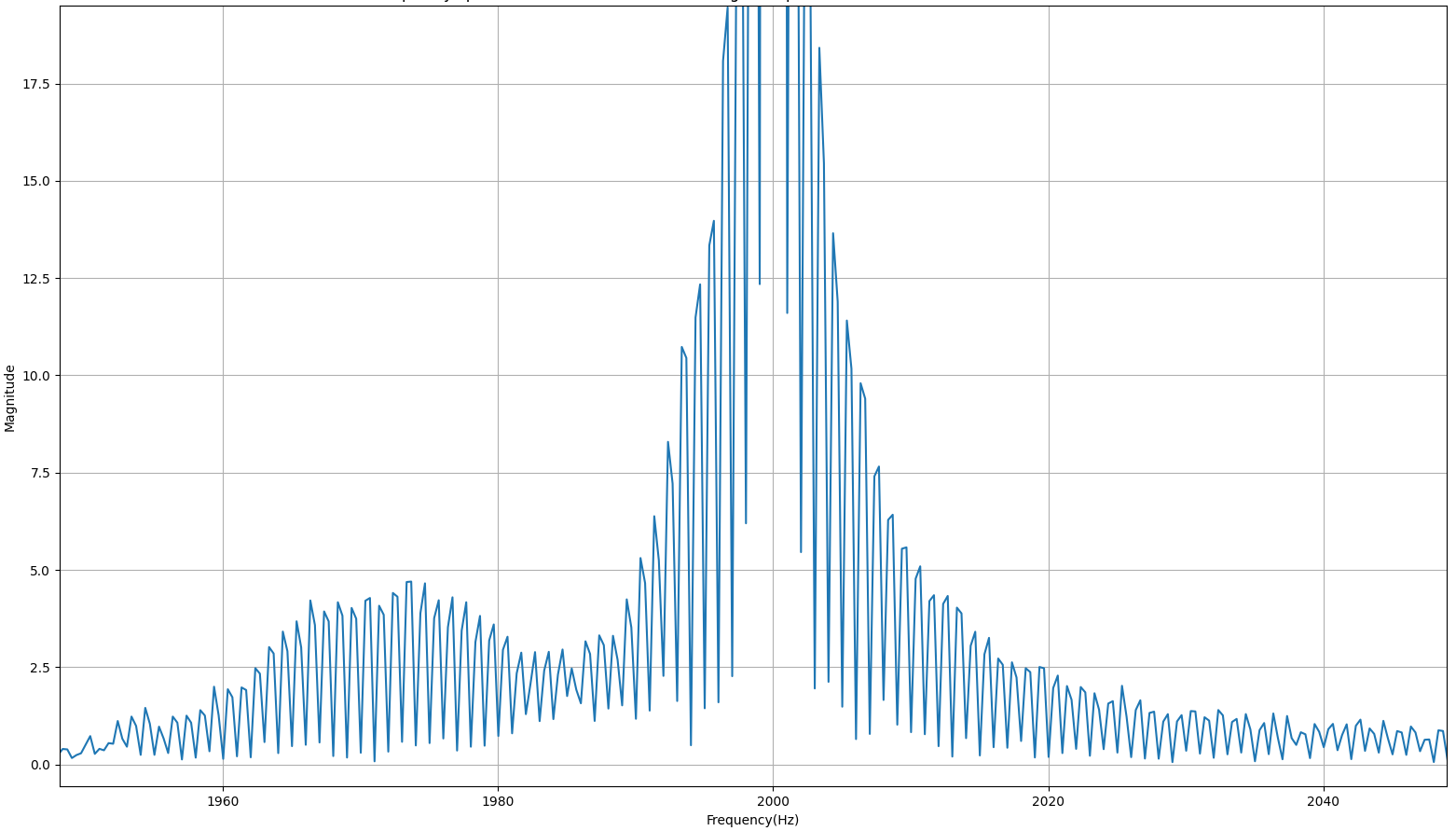}
\includegraphics[width=55mm]{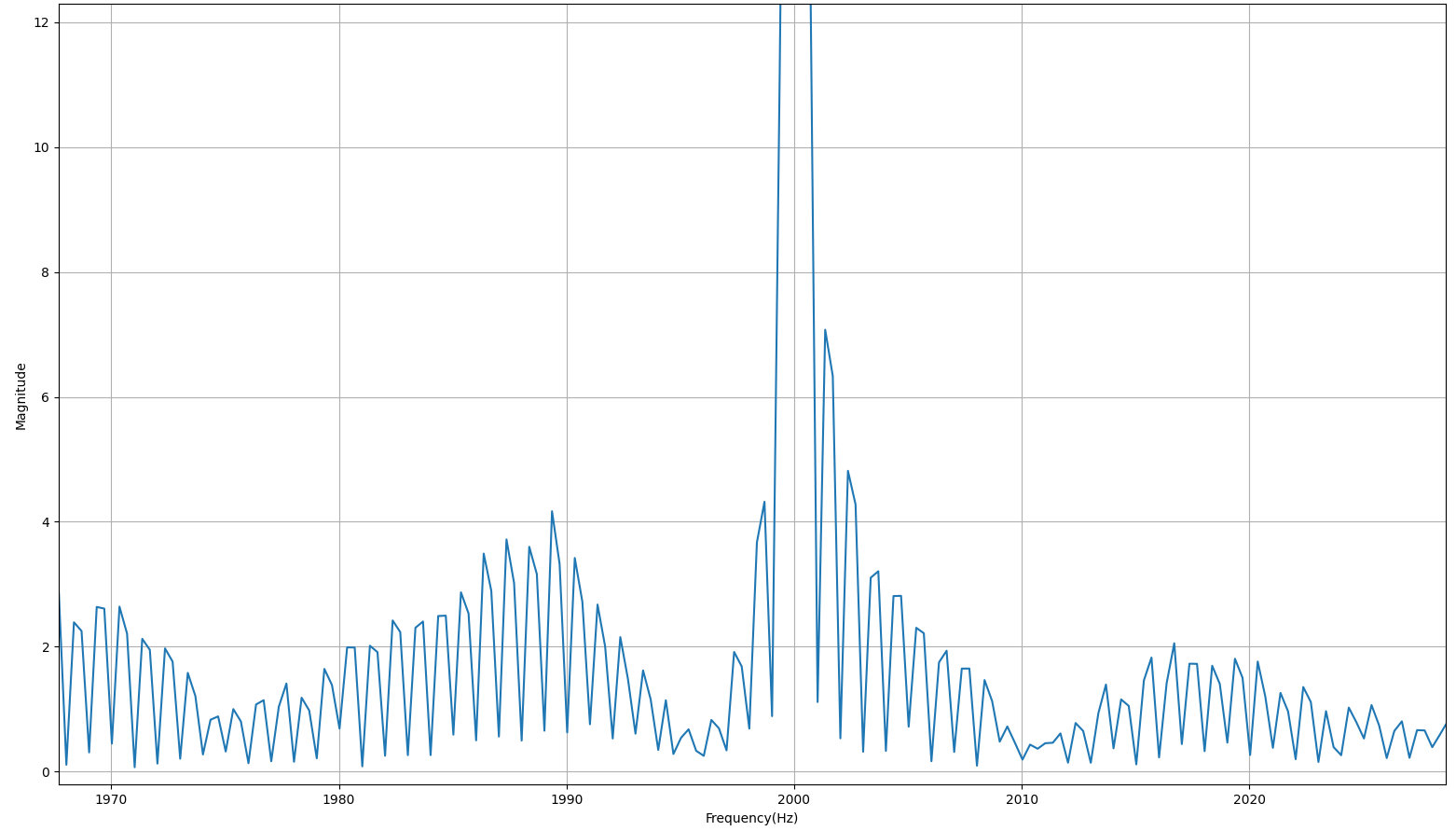}
\includegraphics[width=55mm]{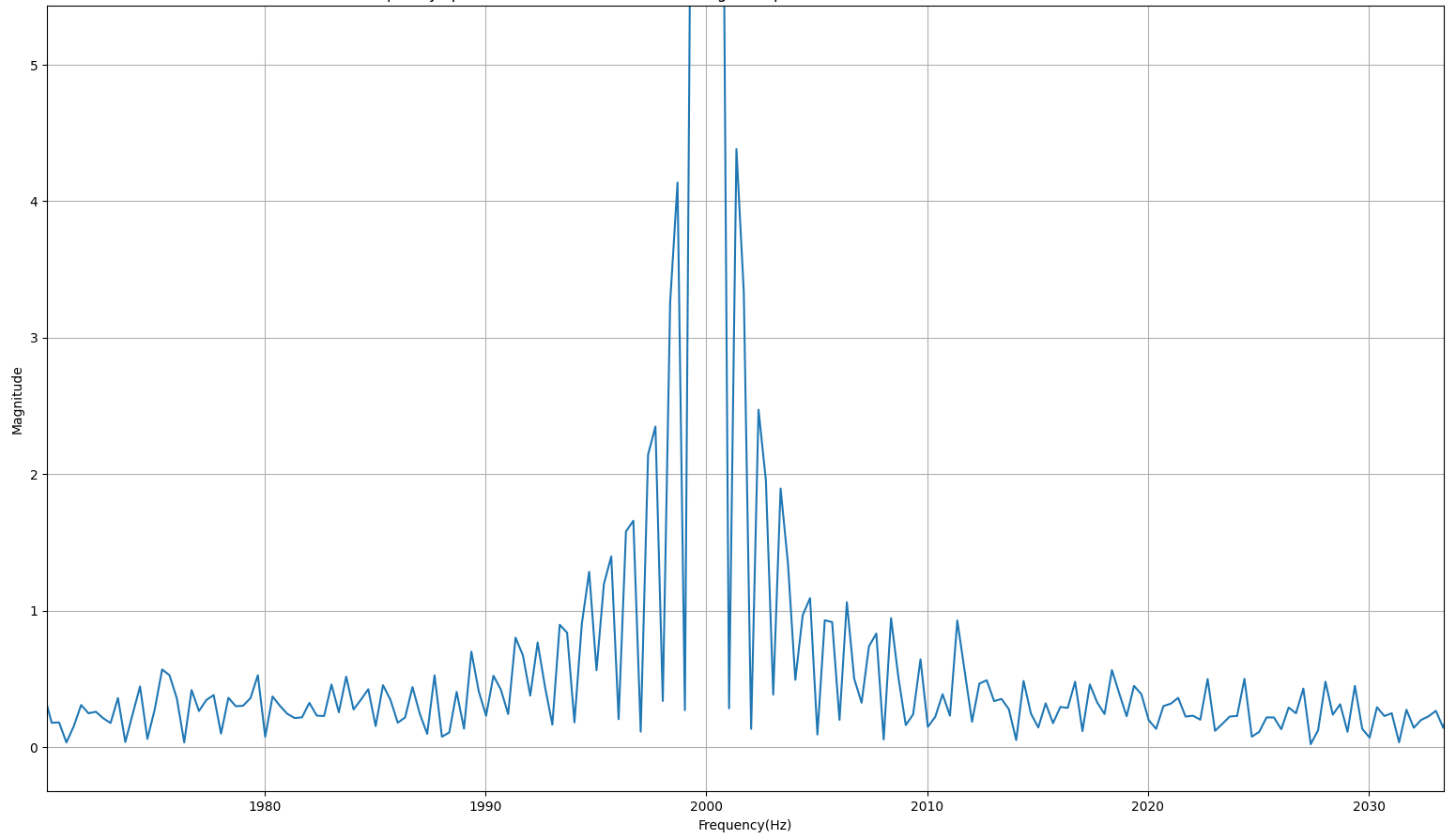}
\caption{Dispersion Patterns of the Frequencies with Most Dispersion at 10' (left), 80' (center), and 161' (right) in NCS.}
\label{fig:MaxDispersion}
\end{figure*}

To determine an upper bound on the information carrying capacity across all frequencies for a pipe run, the theoretical tool of choice is Shannon's Water Pouring Theorem \cite{taub86}.  However, we can also directly estimate the pipe's capacity using the empirical data that we collected to evaluate the Bandwidth Condition for each frequency.  From this data, we know the signal dispersion for each frequency at each distance measured along each pipe run. Given  a set of frequencies that are spaced far enough apart that that they do not interfere with each other according to the dispersion data, the total capacity of that set is the sum of the individual capacities computed according to the Shannon-Hartley Theorem for each frequency.  While this estimate understates the total maximum capacity of the pipe run, it also has an advantage: the lack of interference makes the attacker's task of interpreting the received signal much easier.

Figure \ref{fig:MaxDispersion} shows the dispersion patterns of the frequencies with the \textit{most} dispersion in NCS at 10', 80', and 161'; the patterns in other pipe runs and at other distances were similar. (We do not consider the 4" standpipe in this section, because its diameter is too large for the pipe attack to be effective.) As suggested in the figure,  a frequency separation distance of 80Hz would be needed to ensure that no frequency in the range 15K-20KHz would overlap with any other at any measured distance during transmission.  

We consider a slightly larger separation of 100Hz and consider channels 15K-20KHz in multiples of 100Hz as orthogonal channels that can be used simultaneously. Applying the Shannon-Hartley Theorem to each of the 51 frequencies in the target range and adding up their capacities, we obtain the estimated total capacities shown in the last column of Table \ref{table:individual_channel_capacities}. While a separation distance of 100Hz is overkill, a small transmitter is unlikely to have enough power to transmit on every non-interfering frequency at the same time, unless the transmitter is hardwired or the attack is of quite short duration. Further, parallel transmission on 51 frequencies would be about 15dB louder than the same transmission on a single frequency, which means it would seem roughly three times as loud to a human.  Further, as the number of channels increases, at some point the increase in sound leakage from the pipes might draw human attention.

\graphicspath{{figures/}}
\section{Detection, Localization \& Prevention}
\label{sec:defenses}

We explored the effectiveness of a number of potential defensive measures, none of which so far has proved fully satisfactory. Given the limited space, we omit the details of our implementations and experimental results for these failed defenses.

\textbf{CPVC}.  Though not present in any building we had access to, CPVC is approved as a fire sprinkler pipe material for some building usages in some jurisdictions.  As a non-metallic material, CPVC might be impervious to the pipe attack. However, experiments with air-filled and water-filled .75-1" PVC and CPVC pipe runs in the 30' indoor laboratory setup described earlier showed that in general, frequencies above 500Hz satisfy the Bandwidth Condition for both PVC and CPVC pipes.

\textbf{Check valves.} We  noted in early experiments with radiator water supply loops that we were unable to send signals past a loop's backflow prevention valves. However, experiments in the indoor laboratory with air-filled and water-filled 30'  CPVC and PVC pipes and metal and PVC check values with two different check mechanisms ({Homewerks Worldwide Lead Free Brass FIP x FIP Swing Check Valve, Proline Series FIP x FIP PVC Check Valve})  showed that the check valves had a minuscule effect at best on signal transmission.

  \textbf{Flexible couplings.}
 Victaulic, a major manufacturer of sprinkler system couplings, cites significant reductions in annoying vibrations if three flexible Victaulic couplings are included in a pipe run after the pipe emerges from a pump. We ran experiments  with the  Victaulic 75 flexible and Victaulic 005 rigid grooved couplings, installing three of each model in separate  runs of 42.5' 1.25"  black steel pipe in the outdoor  laboratory. For comparison, we also installed a set of three flexible PVC couplings in a separate run of the lab. In all three runs, single-frequency signals above 400Hz generally satisfied the Bandwidth Condition at all distances tested, sometimes growing stronger than the original transmission along the way.
  
  \textbf{Gaps along the pipe surface.} If signals are traveling primarily through the pipe surface, then the signals might have difficulty in moving past  gaps in the pipe surface. We tested this hypothesis in the outdoor laboratory by adding Everbilt flanged rubber washers to the ends of each segment of pipe before attaching the three models of flexible couplings described above. This only improved signal transmission. 
     
  \textbf{Wider pipes.} Since the pipe attack is less effective in wider sections of pipe, the inclusion of wider sections of pipe at periodic intervals might serve as an effective defense. We have not yet tested this method. We recognize that it would  not be practical in all building locations, because space is often at a premium along sprinkler pipe runs.
  
  \textbf{Ambient noise.}  Ambient noise might reduce the effectiveness of transmissions. However, neither noisy children nor loud white noise played next to the pipe noticeably impaired signal transmission. This suggests that the pipe attack may also be effective in noisy environments like planes, trains, and factories. 
  
    \textbf{Signal  obfuscation.} A defender could inject signals directly into the pipe to obfuscate an attacker's signal. This would increase the cost of a successful attack, but noise injection offers no theoretical guarantees against determined attackers \cite{singlemic_SSL}. It would also add to construction and maintenance costs and might annoy occupants.
  
  \textbf{Sound cancellation.}  A defender could install sound cancellation devices along the pipes.  However, today's noise-cancelling  hardware reacts at the speed of sound in air, which is 331.29 meters per second for dry air at 0 degrees Celsius. Sound travelling as vibrations on a steel pipe surface would be as fast as 3000 meters per second, or possibly higher depending on the composition of the steel, which would pose an extra challenge for real-time response. The price of the resulting system would be so high that it would only be practical for high-stakes environments.

 \textbf{Bug sweeping.} 
 Many bug-sweepers look for RF signals, but this technique will not detect the pipe attack itself, because the attack does not rely on RF. If an attacker  chooses to transfer data to a pipe attack transmitter using RF signals, then  an RF bug sweep could detect that capability if the transmitter is not powered off at the time of the bug sweep.  However, if the transmitter is specially designed to passively watch for RF signals without emitting any itself, then the transmitter itself would not be detected; and it might be hard to distinguish between an ordinary RF-enabled smart light bulb/plug and an attack-enabled bulb/plug that also sends data to the transmitter. The same considerations hold for pipe attack receivers.
 
 To avoid detection by thermal imaging, the transmitter/receiver housing (e.g., faux pipe stub) can be wrapped in mylar or the equivalent to prevent heat radiation from batteries, or direct wired to one of the many power supplies already present above the ceiling. Alternatively, the transmitter/receiver can be passive, i.e., not rely on a power supply at all;  such transmitters were  made and used during the 1950s \cite{TheThing}. 
 
 Non-linear junction detectors (NLJDs) detect the presence of semiconductors even in devices that are powered off, using radiated RF energy to detect harmonic signatures coming off of electronics that contain semiconductor junctions (diodes, transistors, circuit board connections). 

To fool NLJD bug sweepers, the standard countermeasure is to combine the device  with an isolator that will absorb any energy that would otherwise have been reflected back to the NLJD \cite{enwiki:1077008835}.

\textbf{Transmitter localization.}
If a defender believes that a pipe attack is possible, one unappealing option is to visually inspect every inch of the pipe runs until all unauthorized devices have been found and removed. This is made worse by the fact that most building owners do not have blueprints of the sprinkler runs, hence will have to pop up many irrelevant ceiling panels to find them. Further,  defenders  often will not have good visual access in the crowded areas above suspended ceilings, let alone above finished ceilings. 

When a pipe attack transmission is underway, the defender can expedite the search for the transmitter by installing multiple receivers along the pipe run and using the difference in arrival time of the signal at the receivers to determine in which direction the transmitter lies. But this is an insufficient defense on its own, as data can be exfiltrated until the transmitter is found.

 \textbf{Attack recognition.} 
 Given a target false positive/negative rate, a defender can use signal processing techniques to decide whether an attacker's signal is on the pipe, or just noise. But if  the attacker has the technical know-how to keep the signal magnitudes very close to those of the ambient pipe noise, then there is no theoretical guarantee that a defender will be able to recognize that data is being exfiltrated.

\section{Conclusions}
We have shown that building code fire safety requirements have the unintended effect of providing relatively high-speed wide-bandwidth data exfiltration pathways in essentially every room of  commercial and high-rise spaces.  We exploited these pathways by attaching a small transducer that  sends vibrations through pipes to a receiver, at the upper end of human hearing and beyond. In the range of frequencies supported by our equipment (up to 20kHz), signals traveled the maximum distance serviced by a  sprinkler system on a single floor of the large buildings available to us, without significant attenuation. We showed that transmissions using single near-ultrasound frequencies (15K-20KHz) can support up to 300K bps, while also being nearly or completely inaudible to humans.
Transmissions at higher frequencies would no doubt also be effective and inaudible, but even in the 15K-20KHz range, an attacker can use multiple non-interfering frequencies in parallel to create a channel that supports an MP4-equivalent bit rate.

We found that sprinkler pipe transmissions  were quite insensitive to ambient noise, suggesting that the attack may also work in noisy environments like public transportation.
In addition to fire sprinkler system pipes (wet or dry, black steel or CPVC), we showed that the pipe attack is highly effective on potable water and waste pipes (copper, PVC, CPVC) and black steel gas lines. 

The main limitation for the pipe attack appears to be the impracticality of transmitting signals between floors using standpipes. When transmission between floors is required, the best option is to begin the transmission on a convenient sprinkler pipe and then switch over to a nearby pipe of another type at a location such as a restroom, which will offer narrow-diameter risers.
We also considered a wide variety of potential defenses, none of which so far have proven to be both effective and affordable.
 
\label{Defense_eval}

\section*{Acknowledgements}

Special thanks to Laura Kalman, Dustin Lange, Gio Wiederhold, and everyone else who helped us gain access to physical infrastructure in spring and summer 2020, when most large buildings were closed for the pandemic. We also thank Doug L.~Jones and Andrew Singer for their advice and guidance on the audio and communications approaches taken in this paper.


\bibliographystyle{ieeetr}
\bibliography{references}

\end{document}